\documentclass[apj]{emulateapj}

\def\ni{\noindent}

\begin{document}

\shorttitle{Theoretical IRAC Band P-L and P-C Relations}
\shortauthors{Ngeow et al.}

\title{Theoretical Cepheid Period-Luminosity And Period-Color Relations In {\it Spitzer} IRAC Bands}

\author{Chow-Choong Ngeow}
\affil{Graduate Institute of Astronomy, National Central University, Jhongli City, 32001, Taiwan}

\author{Marcella Marconi}
\affil{Osservatorio Astronomico di Capodimonte, Via Moiariello 16, 80131 Napoli, Italy}

\author{Ilaria Musella}
\affil{Osservatorio Astronomico di Capodimonte, Via Moiariello 16, 80131 Napoli, Italy}

\author{Michele Cignoni}
\affil{Department of Astronomy, Bologna University, via Ranzani 1, 40127 Bologna, Italy}

\and 

\author{Shashi M. Kanbur}
\affil{Department of Physics, State University of New York at Oswego, Oswego, NY 13126, USA} 

\begin{abstract}

In this paper the synthetic period-luminosity (P-L) relations in {\it Spitzer's} IRAC bands, based on a series of theoretical pulsation models with varying metal and helium abundance, were investigated. Selected sets of these synthetic P-L relations were compared to the empirical IRAC band P-L relations recently determined from Galactic and Magellanic Clouds Cepheids. For the Galactic case, synthetic P-L relations from model sets with ($Y=0.26$, $Z=0.01$), ($Y=0.26$, $Z=0.02$) and ($Y=0.28$, $Z=0.02$) agree with the empirical Galactic P-L relations derived from the {\it Hubble Space Telescope} parallaxes. For Magellanic Cloud Cepheids, the synthetic P-L relations from model sets with ($Y=0.25$, $Z=0.008$) agree with both of the empirical Large Magellanic Cloud (LMC) and Small Magellanic Cloud (SMC) P-L relations. Analysis of the synthetic P-L relations from all model sets suggested that the IRAC band P-L relations may not be independent of metallicity, as the P-L slopes and intercepts could be affected by the metallicity and/or helium abundance. We also derive the synthetic period-color (P-C) relations in the IRAC bands. Non-vanishing synthetic P-C relations were found for certain combinations of IRAC band filters and metallicity. However, the synthetic P-C relations disagreed with the $[3.6]-[8.0]$ P-C relation recently found for the Galactic Cepheids. The synthetic $[3.6]-[4.5]$ P-C slope from ($Y=0.25$, $Z=0.008$) model set, on the other hand, is in excellent agreement to the empirical LMC P-C counterpart, if a period range of $1.0<\log(P)<1.8$ is adopted.

\end{abstract}


\section{Introduction}

The mid-infrared Cepheid period-luminosity (P-L, also known as the Leavitt Law) relation will be important in the {\it James Webb Space Telescope} ({\it JWST}) era, as it holds the promise of deriving the Hubble constant at the $\sim2\%$ level \citep{fre10,fre11b}. Motivated by this, the {\it Spitzer's} IRAC band ($3.6\mu{\mathrm m}$, $4.5\mu{\mathrm m}$, $5.8\mu{\mathrm m}$ and $8.0\mu{\mathrm m}$) P-L relations were derived for Cepheids in our Galaxy \citep{mar10}, in the Large Magellanic Cloud \citep[LMC,][]{fre08,nge08,mad09,nge09,sco11}, and in the Small Magellanic Cloud \citep[SMC;][]{nge10}.

The slopes of the IRAC band P-L relations are expected to be insensitive to metallicity \citep{fre08,fre10,fre11b}. The empirical slopes derived from a small number of Galactic Cepheids that possess parallax distances \citep{mar10}, and the Magellanic Cloud Cepheids based on the OGLE-III data \citep[see][for more details]{nge09,nge10} are all consistent with each other. However, these slopes do not agree with the slopes derived from Galactic Cepheids that are based on the infrared surface brightness (IRSB) method \citep{mar10} and from a smaller number of LMC Cepheids \citep{mad09}. Therefore, the aim of this paper is to compare these empirical P-L relations to the synthetic IRAC band P-L relations based on a series of theoretical pulsating models at various metallicities, and investigate the sensitivity of IRAC band PL relations to metallicity. 

Brief description of the pulsation models is given in the next section, and the synthetic P-L relations based on these models are presented in Section 3. We compared these synthetic P-L relations to their empirical counterparts and investigated the possible metallicity dependence of the synthetic IRAC band P-L relations in Sections 4 and 5, respectively. In Section 6, synthetic period-color (P-C) relations in the IRAC bands were also derived. Discussion and conclusion are presented in Section 7.

\section{The Pulsational Models}

The synthetic P-L relations adopted in this paper are based on extensive and detailed sets of nonlinear, pulsation models including a non-local, time-dependent treatment of the coupling between pulsation and convection. These models allow us to predict not only the periods and the blue boundary of the instability strip, but also the pulsation amplitudes, the detailed light and radial velocity curve morphology, and the complete topology of the strip, including the red edge \citep{bon99,fio02,mar05,marc10}. For each chemical composition and mass, an evolutionary mass-luminosity (M-L) relation was adopted \citep[see][for details]{mar05} and a wide range of effective temperature was explored. We note that even if the effect of varying the M-L relations has been investigated for specific chemical compositions \citep[see, for examples,][]{bon99,bon00,cap05} in this analysis we assume for all the chemical composition a canonical M-L relation, neglecting both mass loss and overshooting during the previous $H$ burning phase. This is a limitation of the adopted model sets as the above mentioned phenomena affect the P-L relations and the corresponding distance determinations (see Section 7). In total, 17 sets of model with varying helium ($Y$) and metal ($Z$) abundance are considered in this paper, most of them are the same model sets presented in \citet{bon10}.
 
From the resulting theoretical instability strips and the relations connecting the periods to the intrinsic stellar parameters, synthetic P-L relations have been constructed. To this purpose, we populated the predicted instability strip by adopting the procedure suggested by \citet{ken98}. In particular, $\sim 1000$ pulsators were uniformly distributed from the blue to the red boundary of the instability strip, with a mass law as given by $\mathrm{d}n/\mathrm{d}m = m^{-3}$ over the mass range $5$ - $11 M_{\odot}$ \citep[see][for further details]{cap00}. In order to translate the pulsational properties of the investigated Cepheid models in the {\it Spitzer} IRAC bands, we have directly convolved the predicted bolometric light curves with the {\it Spitzer} filter profiles using the general integral equation \citep[see, for example,][]{gir2002}:

\begin{equation}
m_{S_\lambda} = -2.5\,\log\left(\frac{ \int_{\lambda_1}^{\lambda_2}\lambda
f_\lambda S_\lambda \mathrm{d}\lambda }{\int_{\lambda_1}^{\lambda_2}\lambda
f_{\lambda}^0 S_\lambda \mathrm{d}\lambda}\right)
\label{eq_photon}
\end{equation}

\ni where $S_\lambda$ is the IRAC Spectral Response Curve\footnote{Available at {\tt http://ssc.spitzer.caltech.edu/irac/}}, $f_\lambda$ is the stellar flux (that corresponds to model atmospheres of known ($T_{\mathrm{eff}}, [M/H], \log g $)), $f_{\lambda}^0$ is the model spectrum of Vega. Concerning the model atmospheres, we have adopted the homogeneous set of updated ATLAS9 Kurucz model atmospheres and synthetic fluxes (new-ODF models)\footnote{Available at {\tt http://kurucz.harvard.edu/grids.html} or {\tt http://wwwuser.oat.ts.astro.it/castelli/grids.html}}.

\begin{figure}
\plotone{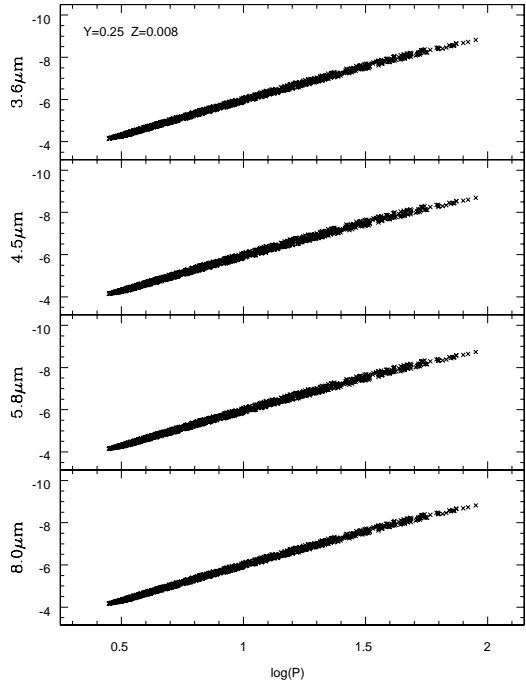}
\caption{Example of the synthetic IRAC band P-L relations, constructed from models with $Y=0.25$ and $Z=0.008$. \label{pl_example}}
\end{figure}

\section{The Synthetic IRAC Band P-L Relations}

\begin{figure*}
\plottwo{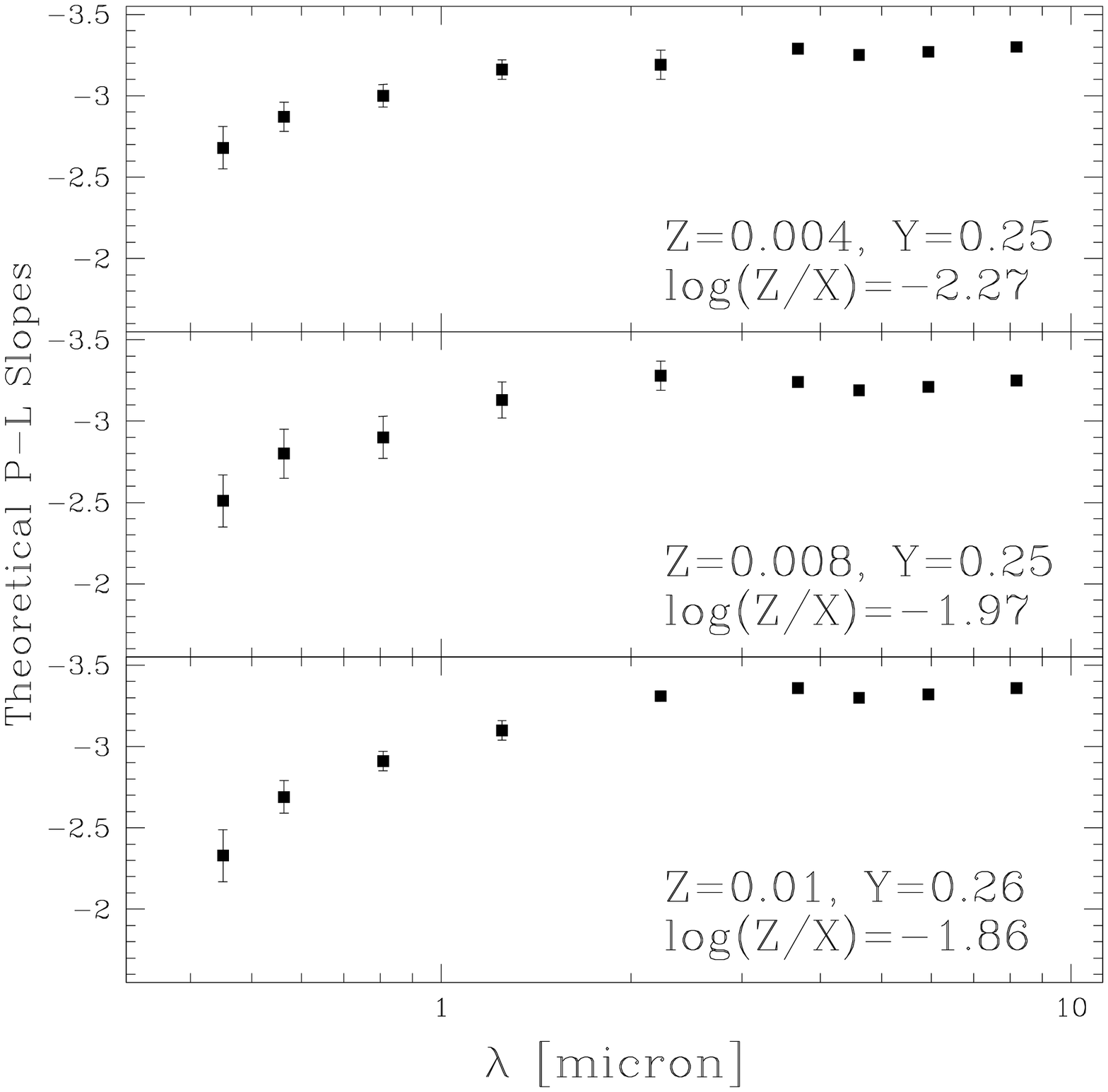}{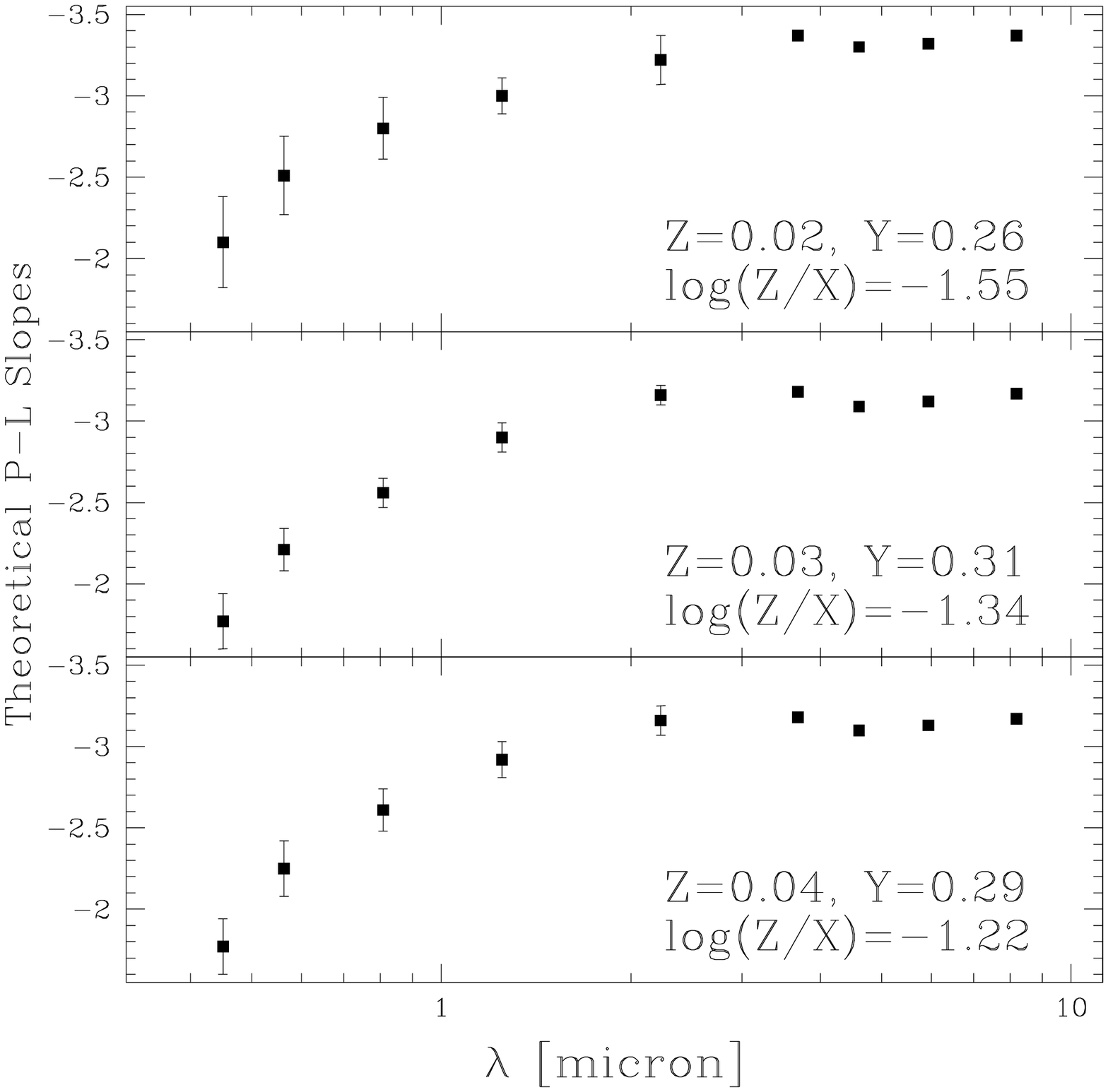}
\caption{Synthetic P-L slopes as a function of wavelength for selected models. Theoretical slopes in $BVIJK$ bands were adopted from \citet[][their Table 2]{bon10}. \label{pl_multiband}}
\end{figure*}

Figure \ref{pl_example} shows an example of the synthetic IRAC band P-L relations from one of the model sets described in the previous section. All the synthetic P-L relations were fitted with the form of $M_{\mathrm{IRAC}}=a+b\times \log(P)$, restricted for pulsators within the period range of $0.4 \leq \log(P) \leq 2.0$ \citep[as in][]{bon10}. The fitted P-L slopes ($b$) and intercepts ($a$) for each of the models were summarized in Table \ref{plslope} and \ref{plzp}, respectively. A majority of the model sets do not have pulsators with $\log(P)<0.4$. For a few of the model sets which have a small number of short-period pulsators, the differences of the P-L slopes and intercepts between the P-L relations derived from full period range and those given in Tables \ref{plslope} and \ref{plzp} do not exceed $0.01$. In Figure \ref{pl_multiband}, the slopes of the synthetic P-L relations in various bands were compared for six of the pulsating model sets (or chemical compositions), where the linear $BVIJK$ P-L slopes were adopted from Table 2 of \citet{bon10}\footnote{For consistency, we only use the linear slopes, $b_{all}$, from \citet{bon10}.}. As expected, the slopes of the P-L relation monotonically decrease from $B$ to $K$ band \citep[see, for example,][]{mad91,ber96,cap00,fio02,fio07,fre08,nge08}, and ``flatten out'' in the mid-infrared. However, the $4.5\mu{\mathrm m}$ P-L relations show a slight increase in their slopes when compared to the ``flatter'' slopes defined from $3.6\mu{\mathrm m}$ and $8.0\mu{\mathrm m}$ band P-L slopes. This slight increase of the $4.5\mu{\mathrm m}$ P-L slopes, and in some extent to the $5.8\mu{\mathrm m}$ band P-L slopes, may be explained due to the presence of CO absorption features shown in the $\sim4\mu{\mathrm m}$ to $\sim6\mu{\mathrm m}$ spectral region \citep[for further details, see][]{mar10,fre11b,sco11}. In fact, the slight increase of $4.5\mu{\mathrm m}$ and $5.8\mu{\mathrm m}$ band P-L slopes was shown in all model sets presented in Table \ref{plslope}.

\begin{deluxetable*}{llcccccccc}
\tabletypesize{\scriptsize}
\tablecaption{Slope of the theoretical IRAC band P-L relations at various metallicities. \label{plslope}}
\tablewidth{0pt}
\tablehead{
\colhead{$Z$} &
\colhead{$Y$} & 
\colhead{$\log(Z/X)$} & 
\colhead{$12+\log(O/H)$\tablenotemark{a}} & 
\colhead{$[\mathrm{Fe}/H]$\tablenotemark{a}} & 
\colhead{$\Delta Y/\Delta Z$\tablenotemark{b}} & 
\colhead{$3.6\mu{\mathrm m}$} & 
\colhead{$4.5\mu{\mathrm m}$} &
\colhead{$5.8\mu{\mathrm m}$} &
\colhead{$8.0\mu{\mathrm m}$} 
}
\startdata
$0.0004$& $0.24$ & $-3.28$ & 7.16 & $-1.50$ & 25 & $-3.511\pm0.004$ & $-3.506\pm0.004$ & $-3.513\pm0.004$ & $-3.519\pm0.004$ \\
$0.001$ & $0.24$ & $-2.87$ & 7.56 & $-1.10$ & 10 & $-3.344\pm0.009$ & $-3.323\pm0.009$ & $-3.336\pm0.009$ & $-3.352\pm0.008$ \\
$0.002$ & $0.24$ & $-2.58$ & 7.86 & $-0.80$ &  5 & $-3.510\pm0.007$ & $-3.485\pm0.007$ & $-3.498\pm0.007$ & $-3.517\pm0.007$ \\
$0.004$ & $0.25$ & $-2.27$ & 8.17 & $-0.49$ &  5 & $-3.292\pm0.005$ & $-3.253\pm0.006$ & $-3.269\pm0.006$ & $-3.298\pm0.005$ \\
$0.006$ & $0.25$ & $-2.09$ & 8.35 & $-0.31$ & 3.3& $-3.365\pm0.007$ & $-3.304\pm0.008$ & $-3.326\pm0.008$ & $-3.371\pm0.007$ \\
$0.008$ & $0.25$ & $-1.97$ & 8.47 & $-0.18$ & 2.5& $-3.244\pm0.005$ & $-3.187\pm0.006$ & $-3.206\pm0.006$ & $-3.248\pm0.005$ \\
$0.01$  & $0.26$ & $-1.86$ & 8.58 & $-0.08$ &  3 & $-3.361\pm0.007$ & $-3.299\pm0.009$ & $-3.318\pm0.008$ & $-3.364\pm0.007$ \\
$0.02$  & $0.25$ & $-1.56$ & 8.88 & $+0.22$ &  1 & $-3.311\pm0.005$ & $-3.239\pm0.006$ & $-3.258\pm0.006$ & $-3.312\pm0.005$ \\
$0.02$  & $0.26$ & $-1.56$ & 8.89 & $+0.23$ & 1.5& $-3.369\pm0.006$ & $-3.304\pm0.007$ & $-3.322\pm0.006$ & $-3.372\pm0.006$ \\
$0.02$  & $0.28$ & $-1.54$ & 8.90 & $+0.24$ & 2.5& $-3.13\pm0.01$\tablenotemark{c} & $-3.04\pm0.01$\tablenotemark{c} & $-3.07\pm0.01$\tablenotemark{c} & $-3.12\pm0.01$\tablenotemark{c} \\
$0.02$  & $0.31$ & $-1.53$ & 8.92 & $+0.26$ &  4 & $-3.271\pm0.005$ & $-3.191\pm0.005$ & $-3.213\pm0.005$ & $-3.272\pm0.005$ \\
$0.03$  & $0.275$& $-1.36$ & 9.08 & $+0.42$ & 1.5& $-3.245\pm0.006$ & $-3.171\pm0.007$ & $-3.191\pm0.006$ & $-3.240\pm0.006$ \\
$0.03$  & $0.31$ & $-1.34$ & 9.10 & $+0.44$ & 2.7& $-3.179\pm0.004$ & $-3.093\pm0.005$ & $-3.118\pm0.004$ & $-3.167\pm0.004$ \\
$0.03$  & $0.335$& $-1.33$ & 9.12 & $+0.46$ & 3.5& $-3.297\pm0.003$ & $-3.210\pm0.004$ & $-3.235\pm0.004$ & $-3.285\pm0.004$ \\
$0.04$  & $0.25$ & $-1.25$ & 9.19 & $+0.53$ & 0.5& $-3.343\pm0.004$ & $-3.268\pm0.005$ & $-3.289\pm0.005$ & $-3.336\pm0.004$ \\
$0.04$  & $0.29$ & $-1.22$ & 9.22 & $+0.56$ & 1.5& $-3.182\pm0.005$ & $-3.104\pm0.005$ & $-3.125\pm0.005$ & $-3.172\pm0.005$ \\
$0.04$  & $0.33$ & $-1.20$ & 9.25 & $+0.59$ & 2.5& $-3.206\pm0.002$ & $-3.129\pm0.003$ & $-3.150\pm0.002$ & $-3.195\pm0.002$  
\enddata
\tablenotetext{a}{Calculated from an online tool, {\tt http://astro.wsu.edu/models/calc/MIX.html}, assuming $12+\log(O/H)_{\odot}=8.66$.}
\tablenotetext{b}{$\Delta Y/\Delta Z = (Y-Y_p)/Z$ is the relative helium enrichment ratio, where $Y_p=0.23$ \citep[see][and reference therein]{fio02}.}
\tablenotetext{c}{Taken from \citet{mar10}.}
\end{deluxetable*}

\begin{deluxetable*}{llcccccccc}
\tabletypesize{\scriptsize}
\tablecaption{Intercept of the theoretical IRAC band P-L relations at various metallicities. \label{plzp}}
\tablewidth{0pt}
\tablehead{
\colhead{$Z$} &
\colhead{$Y$} & 
\colhead{$\log(Z/X)$} & 
\colhead{$12+\log(O/H)$\tablenotemark{a}} & 
\colhead{$[\mathrm{Fe}/H]$\tablenotemark{a}} & 
\colhead{$\Delta Y/\Delta Z$\tablenotemark{b}} & 
\colhead{$3.6\mu{\mathrm m}$} & 
\colhead{$4.5\mu{\mathrm m}$} &
\colhead{$5.8\mu{\mathrm m}$} &
\colhead{$8.0\mu{\mathrm m}$} 
}
\startdata
$0.0004$& $0.24$ & $-3.28$ & 7.16 & $-1.50$ & 25 & $-2.410\pm0.004$ & $-2.418\pm0.004$ & $-2.418\pm0.004$ & $-2.424\pm0.004$ \\
$0.001$ & $0.24$ & $-2.87$ & 7.56 & $-1.10$ & 10 & $-2.527\pm0.010$ & $-2.547\pm0.011$ & $-2.542\pm0.010$ & $-2.539\pm0.010$ \\
$0.002$ & $0.24$ & $-2.58$ & 7.86 & $-0.80$ &  5 & $-2.341\pm0.006$ & $-2.356\pm0.006$ & $-2.353\pm0.006$ & $-2.352\pm0.006$ \\
$0.004$ & $0.25$ & $-2.27$ & 8.17 & $-0.49$ &  5 & $-2.718\pm0.005$ & $-2.740\pm0.006$ & $-2.737\pm0.006$ & $-2.730\pm0.005$ \\
$0.006$ & $0.25$ & $-2.09$ & 8.35 & $-0.31$ & 3.3& $-2.588\pm0.007$ & $-2.621\pm0.008$ & $-2.615\pm0.007$ & $-2.599\pm0.007$ \\
$0.008$ & $0.25$ & $-1.97$ & 8.47 & $-0.18$ & 2.5& $-2.702\pm0.005$ & $-2.724\pm0.006$ & $-2.723\pm0.006$ & $-2.714\pm0.005$ \\
$0.01$  & $0.26$ & $-1.86$ & 8.58 & $-0.08$ &  3 & $-2.577\pm0.007$ & $-2.594\pm0.008$ & $-2.595\pm0.008$ & $-2.590\pm0.007$ \\
$0.02$  & $0.25$ & $-1.56$ & 8.88 & $+0.22$ &  1 & $-2.608\pm0.005$ & $-2.609\pm0.005$ & $-2.617\pm0.005$ & $-2.621\pm0.005$ \\
$0.02$  & $0.26$ & $-1.56$ & 8.89 & $+0.23$ & 1.5& $-2.596\pm0.005$ & $-2.598\pm0.006$ & $-2.605\pm0.006$ & $-2.607\pm0.005$ \\
$0.02$  & $0.28$ & $-1.54$ & 8.90 & $+0.24$ & 2.5& $-2.66\pm0.01$\tablenotemark{c} & $-2.67\pm0.01$\tablenotemark{c} & $-2.69\pm0.01$\tablenotemark{c} & $-2.68\pm0.01$\tablenotemark{c} \\
$0.02$  & $0.31$ & $-1.53$ & 8.92 & $+0.26$ &  4 & $-2.608\pm0.004$ & $-2.627\pm0.005$ & $-2.629\pm0.005$ & $-2.620\pm0.005$ \\
$0.03$  & $0.275$& $-1.36$ & 9.08 & $+0.42$ & 1.5& $-2.630\pm0.006$ & $-2.619\pm0.006$ & $-2.629\pm0.006$ & $-2.647\pm0.006$ \\
$0.03$  & $0.31$ & $-1.34$ & 9.10 & $+0.44$ & 2.7& $-2.637\pm0.004$ & $-2.639\pm0.005$ & $-2.644\pm0.004$ & $-2.660\pm0.004$ \\
$0.03$  & $0.335$& $-1.33$ & 9.12 & $+0.46$ & 3.5& $-2.508\pm0.004$ & $-2.514\pm0.004$ & $-2.518\pm0.004$ & $-2.532\pm0.004$ \\
$0.04$  & $0.25$ & $-1.25$ & 9.19 & $+0.53$ & 0.5& $-2.545\pm0.004$ & $-2.525\pm0.004$ & $-2.537\pm0.004$ & $-2.563\pm0.004$ \\
$0.04$  & $0.29$ & $-1.22$ & 9.22 & $+0.56$ & 1.5& $-2.695\pm0.004$ & $-2.680\pm0.005$ & $-2.690\pm0.005$ & $-2.716\pm0.004$ \\
$0.04$  & $0.33$ & $-1.20$ & 9.25 & $+0.59$ & 2.5& $-2.592\pm0.002$ & $-2.580\pm0.003$ & $-2.589\pm0.002$ & $-2.615\pm0.002$  
\enddata
\tablenotetext{a}{Calculated from an online tool, {\tt http://astro.wsu.edu/models/calc/MIX.html}, assuming $12+\log(O/H)_{\odot}=8.66$.}
\tablenotetext{b}{$\Delta Y/\Delta Z = (Y-Y_p)/Z$ is the relative helium enrichment ratio, where $Y_p=0.23$ \citep[see][and reference therein]{fio02}.}
\tablenotetext{c}{Taken from \citet{mar10}.}
\end{deluxetable*}

\section{Comparison to the Empirical P-L Relations}

Slopes of the current empirical IRAC band P-L relations are summarized in Table 2 of \citet{nge10}. These include six sets of P-L relations in our Galaxy and Magellanic Clouds. Briefly, GAL1 and GAL2 are P-L slopes derived from the ``old'' and ``new'' IRSB distances, respectively, and GAL3 are slopes derived from eight Cepheids that possess {\it Hubble Space Telescope (HST)} parallax measurements. Details concerning these Galactic P-L slopes can be found in \citet{mar10}. The LMC1 and LMC2 are the empirical LMC P-L slopes taken from \citet{mad09} and \citet{nge09}, respectively, and the SMC P-L slopes were adopted from \citet{nge10}. As discussed in \citet{nge10}, these six sets of P-L slopes can be grouped to two groups characterized by steeper ($\sim -3.46$) and shallower ($\sim -3.18$) slopes respectively. Both of the steeper and shallower slopes can be predicted from using $L_{\lambda}=4\pi R^2 B_{\lambda}(T)$, by assuming the behavior of $B_{\lambda}(T)$ at long wavelengths for the IRAC bands \citep[see][for more details]{fre08,nei10,nge10b}. 

In addition to these six sets of empirical P-L relations obtained from random-phase observations, \citet{sco11} have recently published LMC P-L relations based on $\sim 80$ Cepheids that possess mean magnitudes in $3.6\mu{\mathrm m}$ and $4.5\mu{\mathrm m}$ bands. These LMC Cepheids have been observed multiple times using {\it Spitzer}, with $24$ evenly spaced data points per light curves, hence accurate mean magnitudes can be obtained. Their adopted P-L relations in $3.6\mu{\mathrm m}$ and $4.5\mu{\mathrm m}$ bands are denoted as LMC3 in this paper.

These empirical P-L relations can be compared to the synthetic IRAC band P-L relations from the selected model sets given in Tables \ref{plslope} and \ref{plzp}. \citet{bon10} have compared the synthetic $BVIJK$ band P-L slopes of these selected model sets to their empirical counterparts for Galactic and Magellanic Clouds Cepheids, and found they are generally in agreement. When comparing the P-L intercepts, three different values of LMC and SMC distance moduli ($\mu_{\mathrm{LMC,SMC}}$),\footnote{Extinction is ignored as it is negligible in IRAC bands \citep{fre08,fre10,nge09}.} respectively, were adopted (see the right panels of Figures \ref{pl_compare_lmc} and \ref{pl_compare_smc}). These distance moduli covered a wide range of available distance moduli in the literature for the Magellanic Clouds.

\subsection{Comparison of the Galactic P-L Relations}

\begin{figure*}
\plottwo{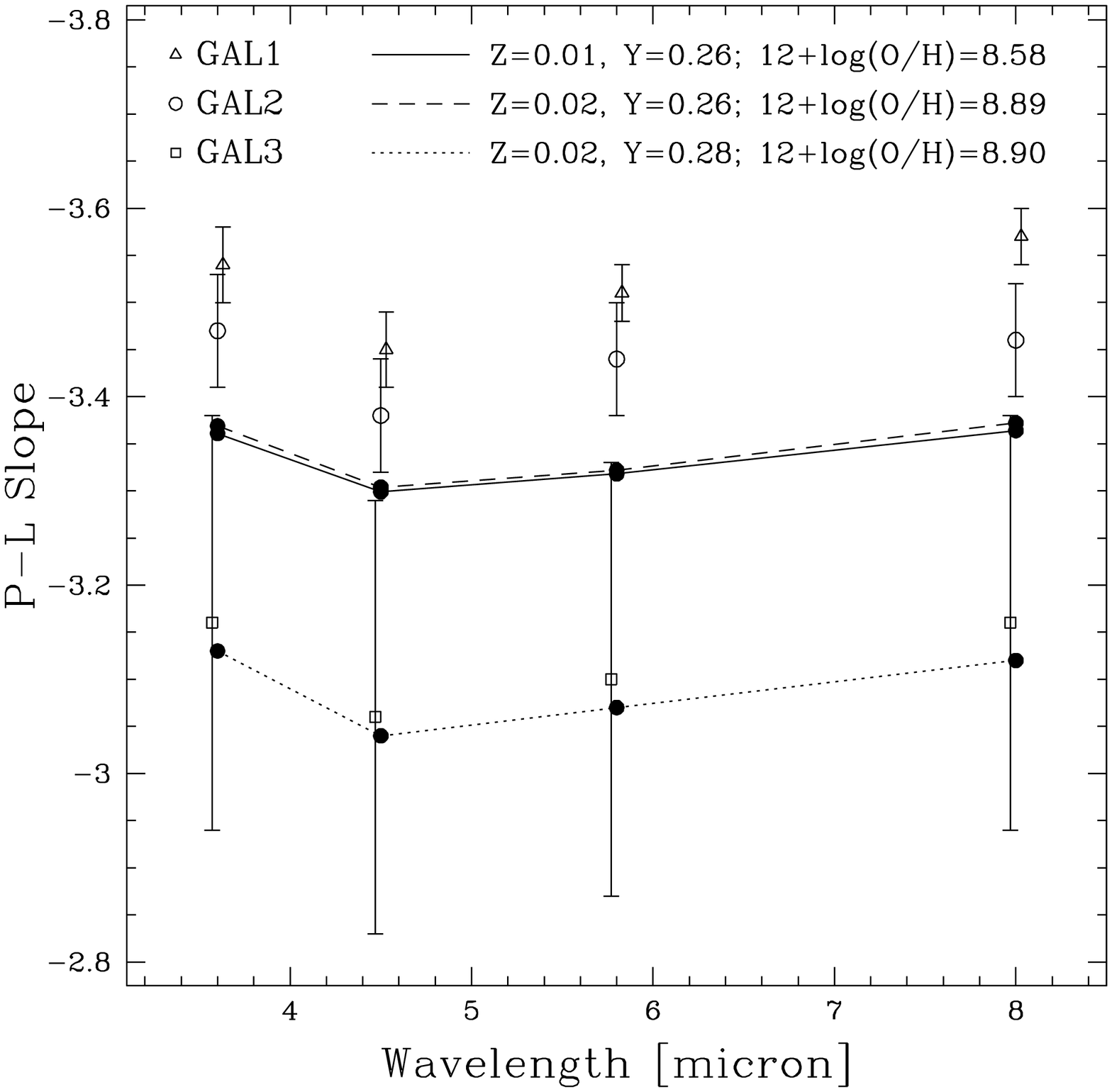}{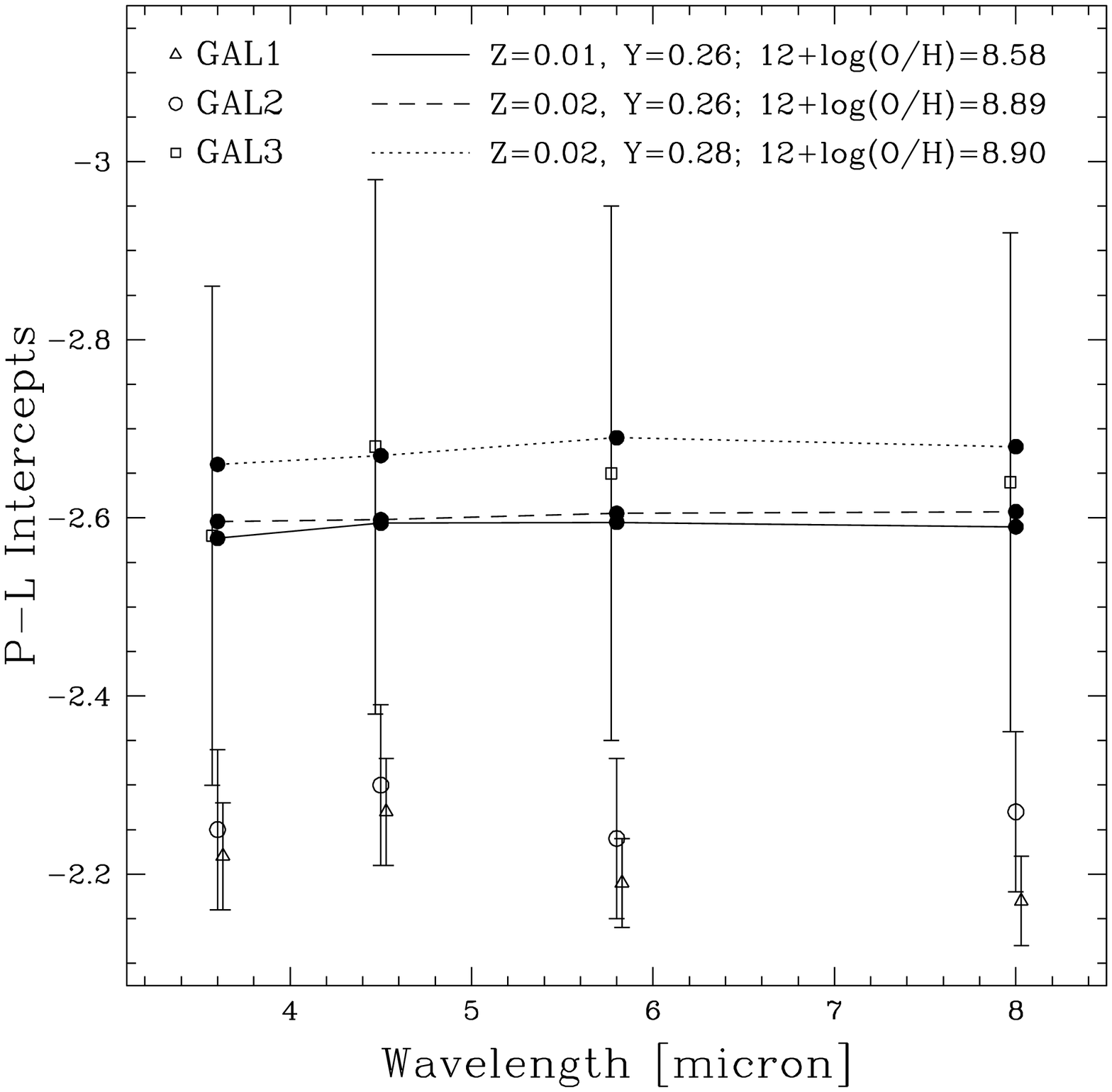}
\caption{Comparison of the empirical Galactic P-L relations, adopted from \citet{mar10}, to the selected synthetic P-L relations given in Table \ref{plslope} (left panel) and \ref{plzp} (right panel). Note that for better visualization, data points for GAL1 and GAL3 have been shifted slightly in wavelength. See the text for the definition of GAL1, GAL2 and GAL3. \label{pl_compare_gal}}
\end{figure*} 

A comparison of the synthetic and empirical Galactic P-L relations in IRAC bands are presented in Figure \ref{pl_compare_gal}. The left panel of this figure shows that the synthetic P-L slopes from $12+\log(O/H)=8.58$ ($Z=0.01$, $Y=0.26$) and $12+\log(O/H)=8.89$ ($Z=0.02$, $Y=0.26$) model sets are in marginal agreement with the GAL2 and GAL3 P-L slopes, but not for the GAL1 P-L slopes derived from the ``old'' IRSB distances. However, the synthetic P-L intercepts from these two model sets agree well with the GAL3 P-L intercepts (albeit the large error bars), but disagree with the other two empirical P-L intercepts based on the IRSB methods. \citet{mar10} have compared the synthetic P-L relations from $12+\log(O/H)=8.90$ ($Z=0.02$, $Y=0.28$) model set to the three sets of empirical Galactic P-L relations. They are also included in Figure \ref{pl_compare_gal}. The comparison shown in this figure echoes the finding in \citet{mar10} that this set of synthetic P-L relations agree well with their empirical counterparts derived from the {\it HST} parallaxes (GAL3), but not for the other two sets of empirical Galactic P-L relations. 

Determination of the Galactic P-L relations is expected to improve in the near future. The Carnegie Hubble Program will observe and measure the distances to about 39 Galactic Cepheids in $3.6\mu{\mathrm m}$ and $4.5\mu{\mathrm m}$, with 15 of them expected to have parallax measurements from {\it Gaia} \citep{fre10}. The improved empirical Galactic P-L relations are expected to be able to discriminate the synthetic P-L relations that are best at describing the observed P-L relations.

\subsection{Comparison of the LMC P-L Relations}

\begin{figure*}
\plottwo{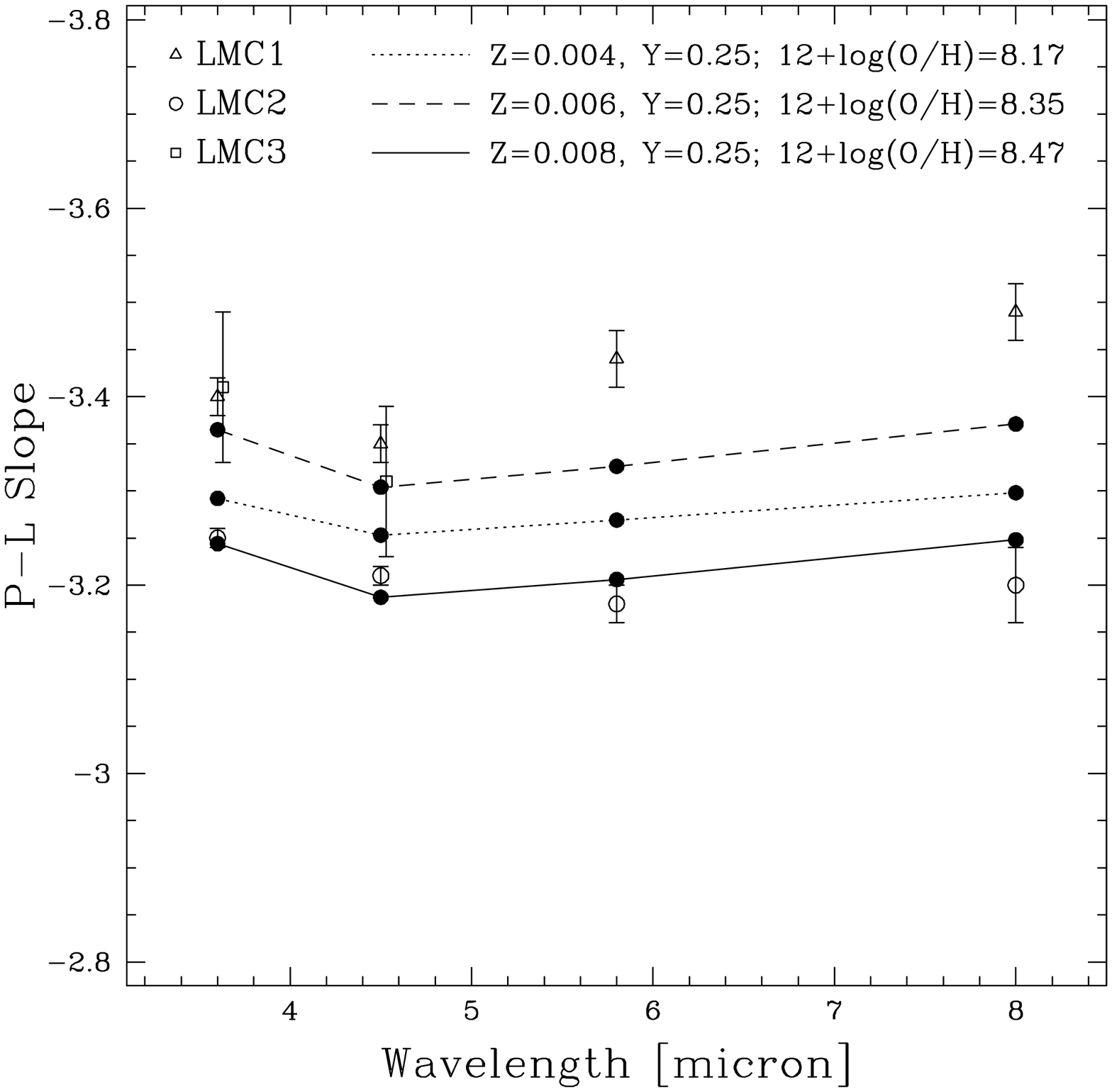}{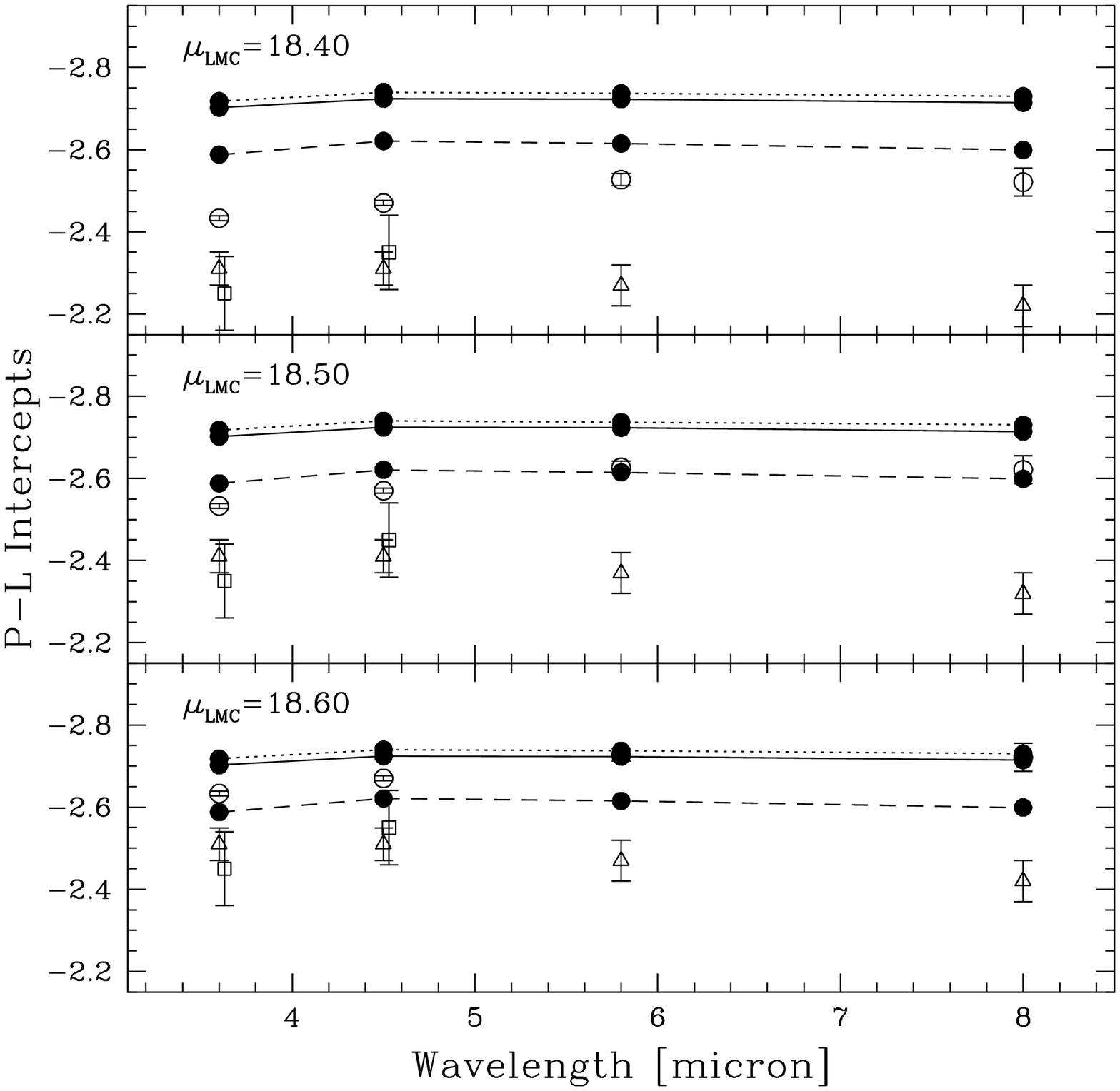}
\caption{Same as Figure \ref{pl_compare_gal}, assuming three different values of the LMC distance modulus. These distance moduli roughly cover the available LMC distance modulus in literature. Note that for better visualization, data points for LMC3 have been shifted slightly in wavelength.\label{pl_compare_lmc}}
\end{figure*}

In Figure \ref{pl_compare_lmc}, synthetic P-L relations from three model sets were compared to the empirical LMC P-L relations. The synthetic P-L slopes from $12+\log(O/H)=8.47$ ($Z=0.008$, $Y=0.25$) model set are in good agreement with LMC2 P-L slopes from \citet{nge09}. Note that the model set with $Z=0.008$ and $Y=0.25$ is generally adopted as a representative metallicity for LMC Cepheids. On the other hand, the synthetic P-L slopes from the $12+\log(O/H)=8.35$ ($Z=0.006$, $Y=0.25$) model set also agree with LMC1 P-L slopes \citep[from][]{mad09} in $3.6\mu{\mathrm m}$ and $4.5\mu{\mathrm m}$ bands, but not in the two longer wavelength bands. Similarly, LMC3 P-L slopes from \citet{sco11} agreed with the synthetic P-L slopes from the same model set. For the P-L intercepts, the right panels of Figure \ref{pl_compare_lmc} show that the synthetic P-L intercepts from $12+\log(O/H)=8.35$ model set match with the empirical results from LMC2 if the LMC distance modulus ($\mu_{\mathrm LMC}$) is adopted to be $18.50$ mag. The LMC2 P-L intercepts at $5.8\mu{\mathrm m}$ and $8.0\mu{\mathrm m}$ bands also matched to the synthetic P-L intercepts from $12+\log(O/H)=8.47$ and $12+\log(O/H)=8.17$ model sets if $\mu_{{\mathrm LMC}}\sim 18.60$ mag.

The P-L slopes adopted from \citet{sco11} are based on Cepheids with period range of $1.0<\log (P)<1.8$. Table 3 of \citet{sco11} also listed the P-L relations using Cepheids with $0.8<\log (P)<1.8$, at which the slopes of these P-L relations ($-3.31\pm0.05$ and $-3.22\pm0.05$ at $3.6\mu{\mathrm m}$ and $4.5\mu{\mathrm m}$ bands, respectively) are in better agreement to the LMC2 empirical P-L slopes from \citet{nge09} and the synthetic P-L slopes from $Z=0.004$ and $Z=0.008$ (both with $Y=0.25$) model sets given in tje previous section. The synthetic P-L slopes using restricted period ranges, either for $1.0<\log (P)<1.8$ or $0.8<\log (P)<1.8$, however, do not reproduce the steep slopes as adopted in \citet{sco11}. In contrast, these P-L slopes are shallower than the synthetic P-L slopes from the full period range ($0.4<\log [P]< 2.0$). For examples, $3.6\mu{\mathrm m}$ band P-L slopes for $Z=(0.004,\ 0.006,\ 0.008)$ model sets with $1.0<\log (P)<1.8$ are $-3.153\pm0.019$, $-3.223\pm0.022$ and $-3.134\pm0.017$, respectively. Similarly, the $4.5\mu{\mathrm m}$ band P-L slopes for $Z=(0.004,\ 0.006,\ 0.008)$ model sets are $-3.081\pm0.022$, $-3.125\pm0.025$ and $-3.046\pm0.021$, respectively. This suggests that the adopted period range could affect the derived P-L relations. Interestingly, \citet{nei10} demonstrated that the empirical P-L slopes based on \citet{nge09} data can be steepen and consistent to the P-L slopes from \citet{mad09}, or \citet{sco11}, if a period cut of $\log(P)=1.05$ is applied.

\subsection{Comparison of the SMC P-L Relations}

\begin{figure*}
\plottwo{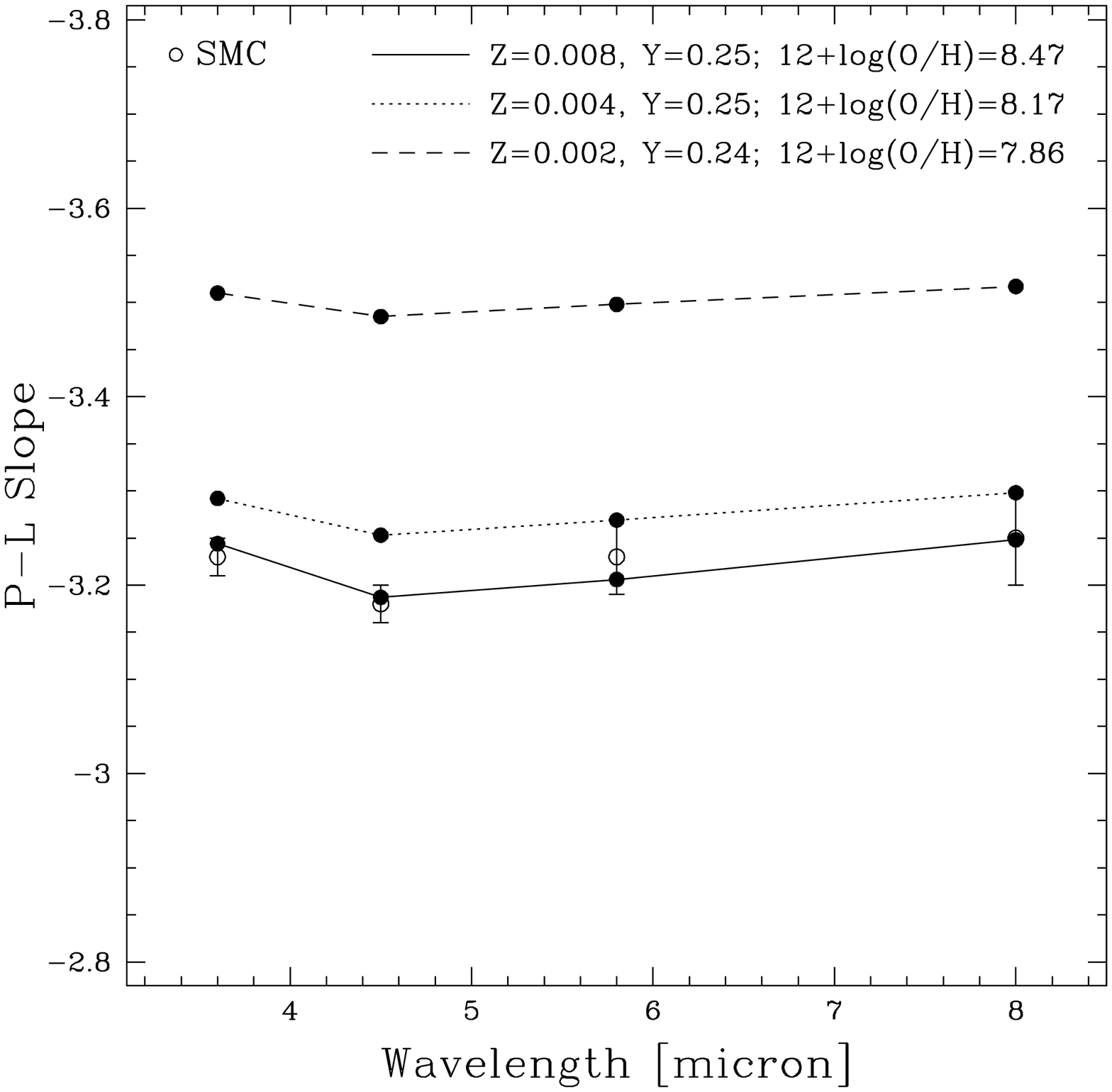}{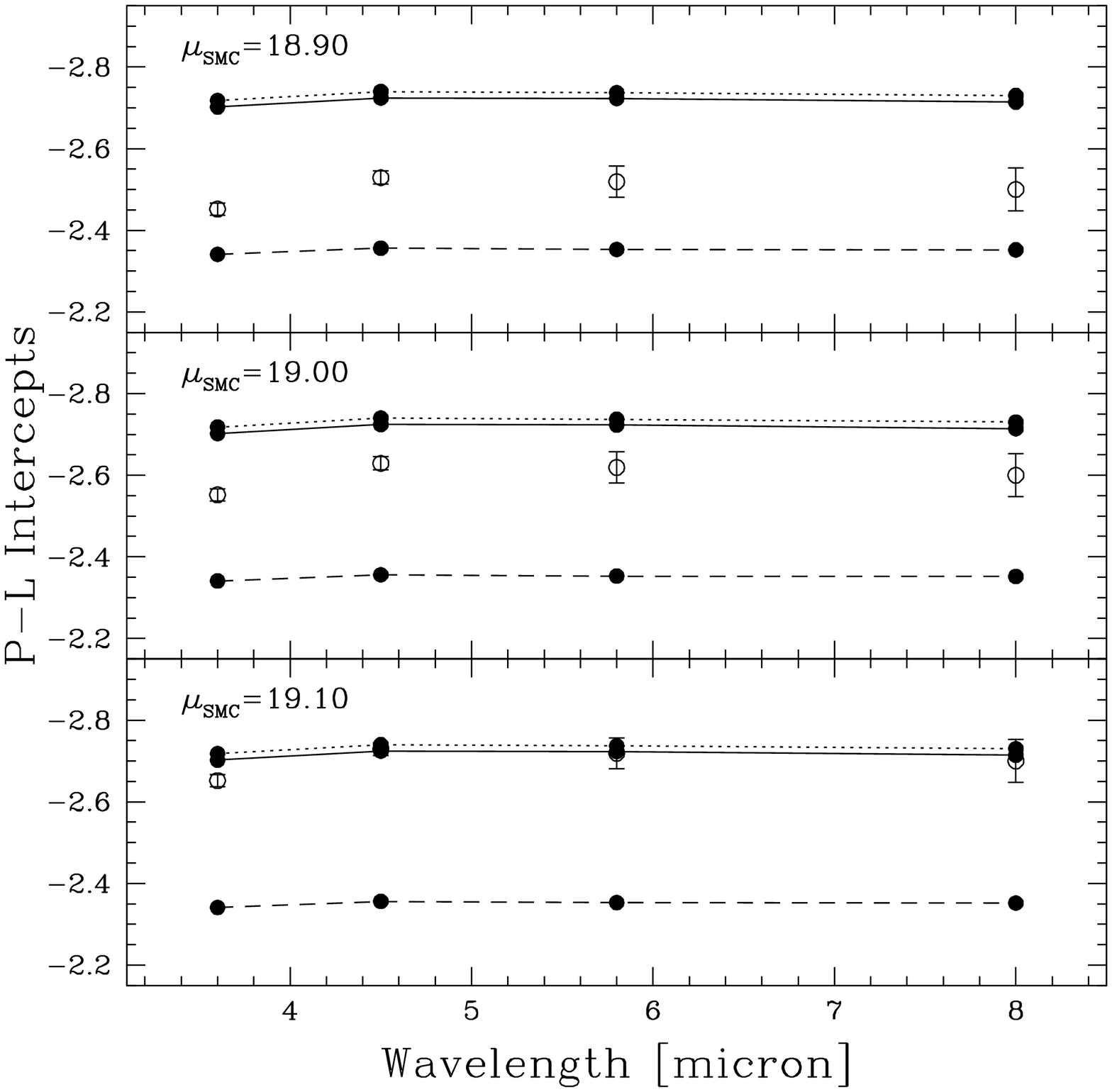}
\caption{Same as Figure \ref{pl_compare_gal}, assuming three different values of the SMC distance modulus. These distance moduli roughly cover the available SMC distance modulus in literature. \label{pl_compare_smc}}
\end{figure*}

Synthetic P-L relations from three model sets, $12+\log(O/H)=7.86$ ($Z=0.002$, $Y=0.24$), $12+\log(O/H)=8.17$ ($Z=0.004$, $Y=0.25$) and $12+\log(O/H)=8.47$ ($Z=0.008$, $Y=0.25$), were compared to the empirical SMC P-L relations from \citet{nge10}. As can be seen from Figure \ref{pl_compare_smc}, the synthetic P-L relations from $12+\log(O/H)=8.47$ model set agree with the empirical P-L relations, if the assumed SMC distance modulus is $19.10$ mag. This result is odd because $Z=0.004$ is generally adopted as representative of SMC metallicity. Though there is a spread in the metallicity of SMC Cepheids from spectroscopic measurements, the mean value is closer to $Z=0.004$ than to $Z=0.008$. This occurrence is due to the fact that current nonlinear pulsation models predict a significant steepening of P-L slopes when changing the metallicity from $Z=0.008$ to $Z=0.004$ (with a significant reduction of the dependence at still smaller metal contents) whereas the empirical relations adopted in this paper tend to suggest almost the same slopes for LMC and SMC. For $12+\log(O/H)=8.17$ model set, the synthetic P-L slopes agree with the empirical P-L relations at $5.8\mu{\mathrm m}$ and $8.0\mu{\mathrm m}$ bands, though the $3.6\mu{\mathrm m}$ and $4.5\mu{\mathrm m}$ P-L slopes are slightly off from the empirical counterparts.

\section{Dependency of Theoretical P-L Relations on Metallicity}

\begin{deluxetable}{lcc}
\tabletypesize{\scriptsize}
\tablecaption{$F$-test results for synthetic P-L relations as a function of metallicity. \label{ftest1}}
\tablewidth{0pt}
\tablehead{
\colhead{Band} &
\colhead{P-L Slope} & 
\colhead{P-L Intercept}  
}
\startdata
$3.6\mu{\mathrm m}$ & 15.3 & 5.68 \\
$4.5\mu{\mathrm m}$ & 24.3 & 3.47 \\
$5.8\mu{\mathrm m}$ & 23.3 & 4.34 \\
$8.0\mu{\mathrm m}$ & 17.2 & 6.46 
\enddata
\tablecomments{We only list the results for $12+\log(O/H)$. The $F$-test results for $[\mathrm{Fe}/H]$ are very similar and hence not listed in this Table.}
\end{deluxetable}

\begin{figure*}
\plottwo{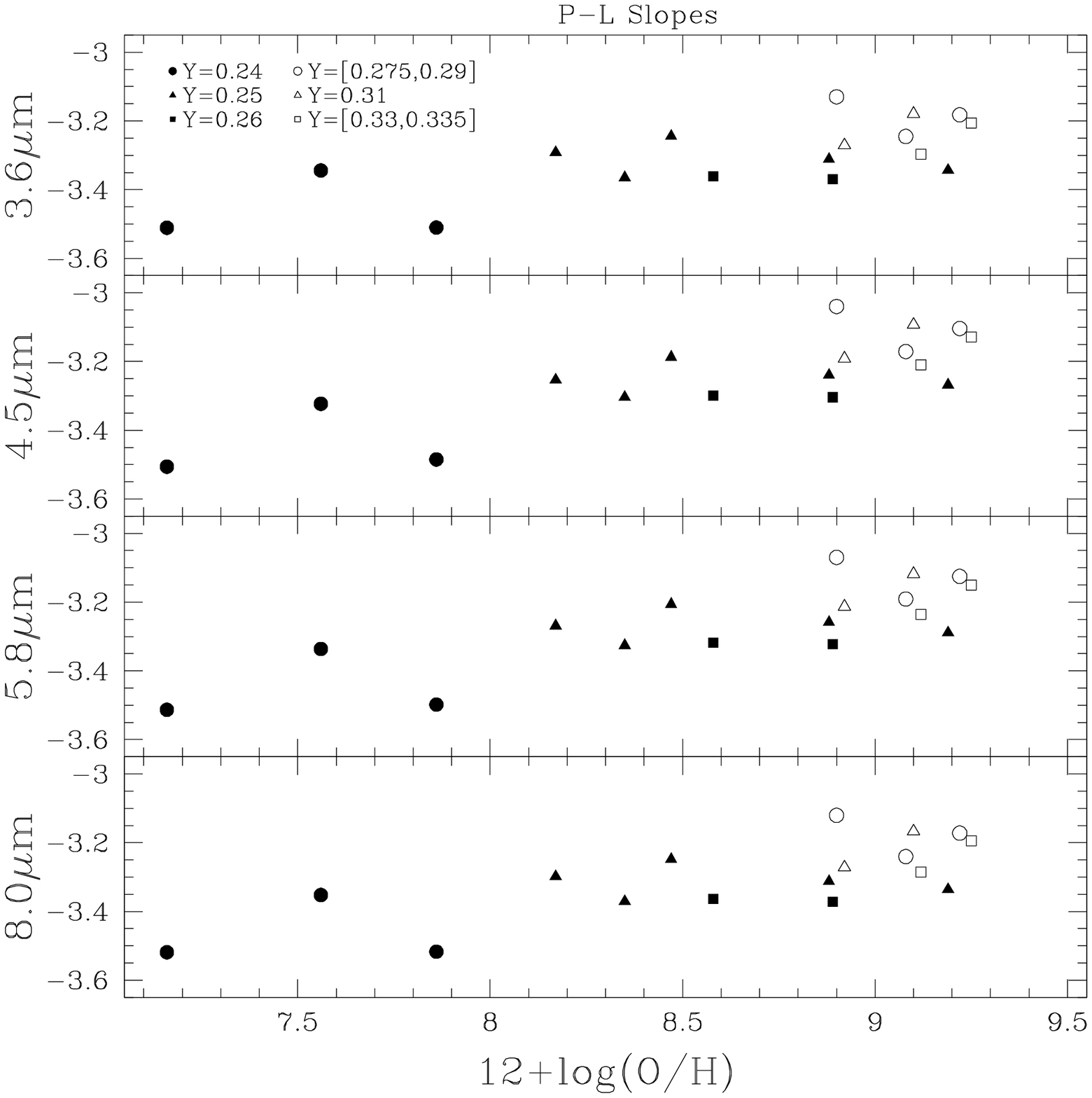}{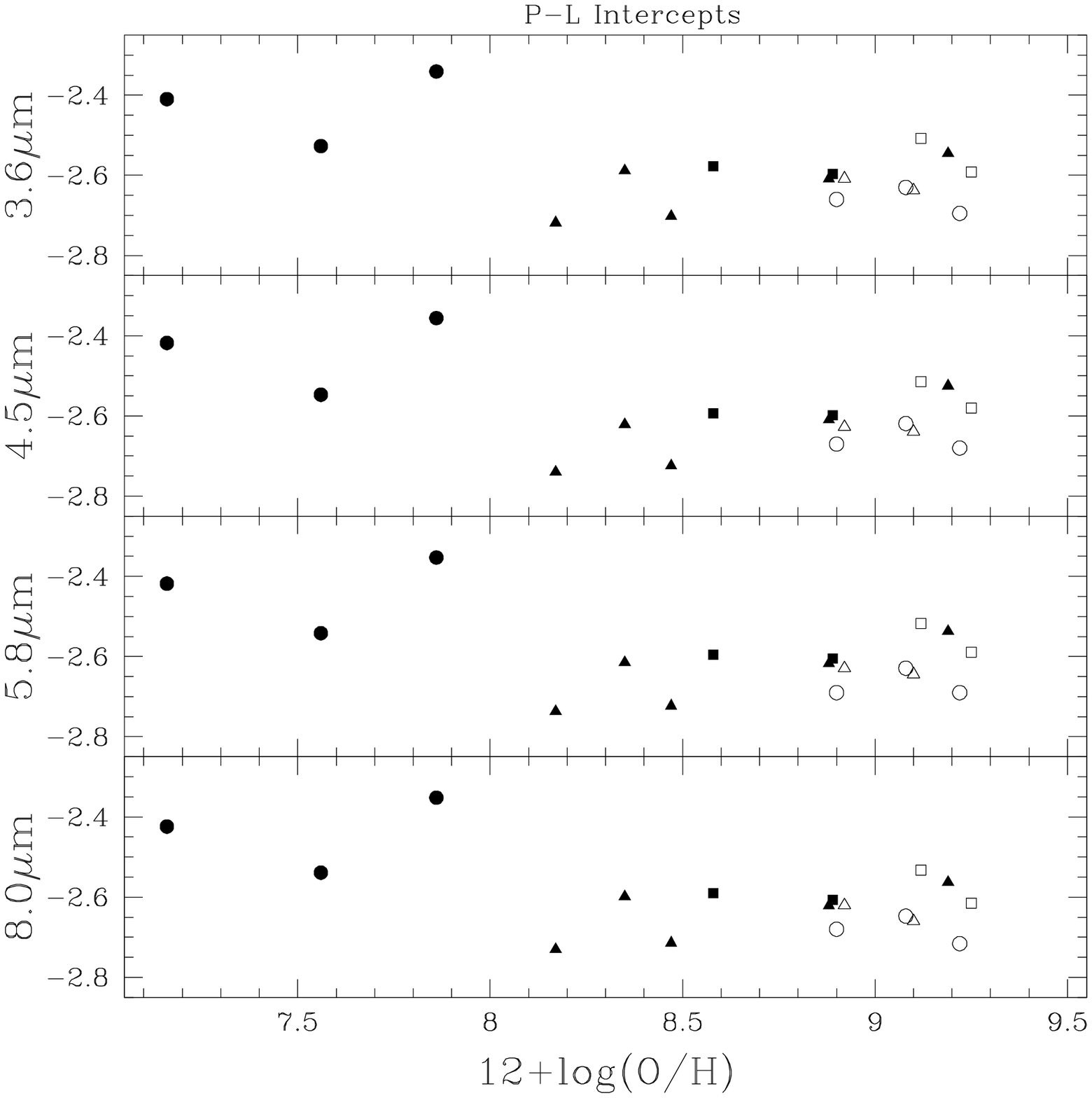}
\caption{Synthetic P-L relations as a function of $12+\log(O/H)$, separated by the helium abundance ($Y$). The left and right panels are for the P-L slopes and intercepts, respectively. \label{pl_zx}}
\end{figure*} 

\begin{figure*}
\plottwo{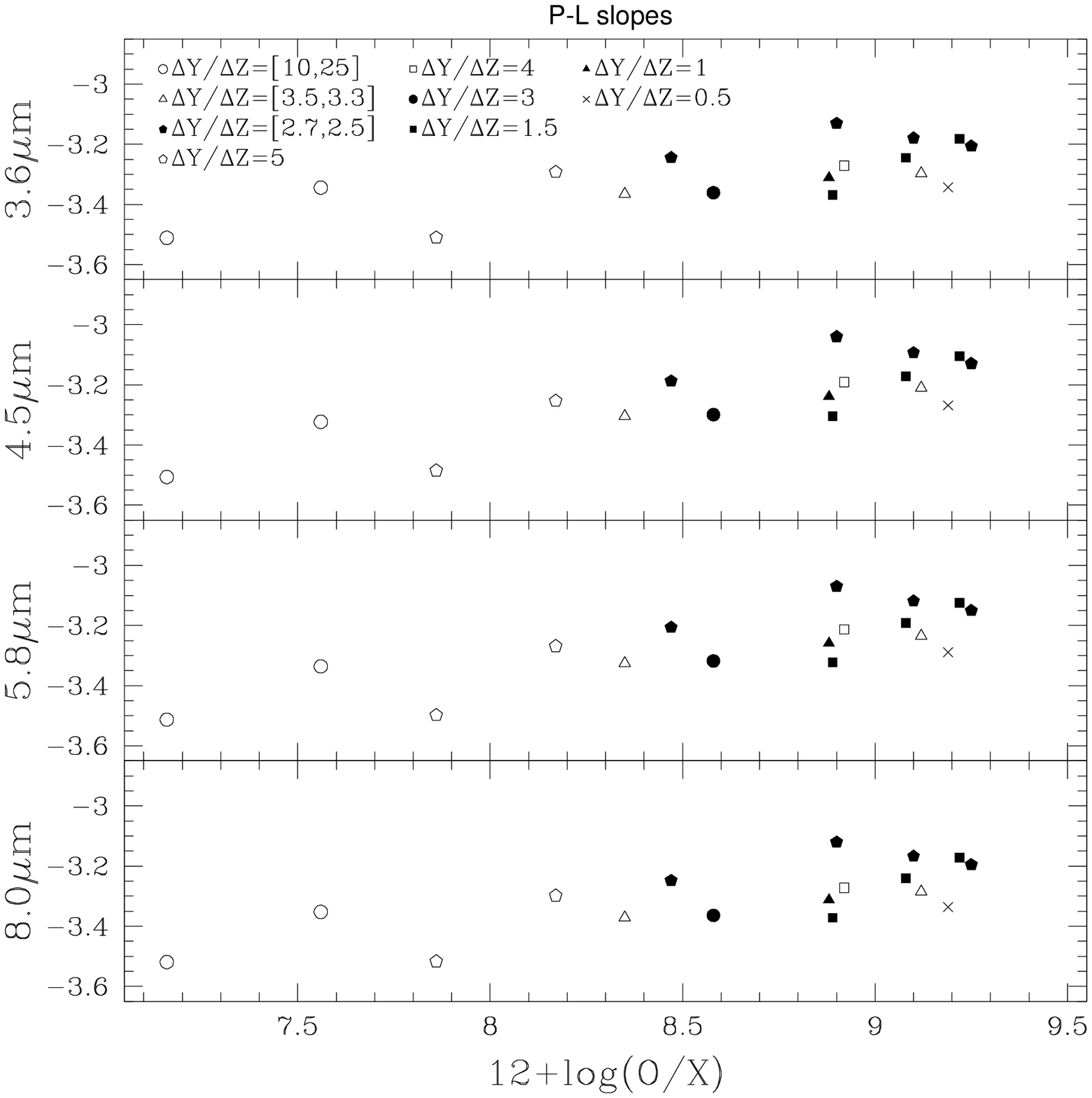}{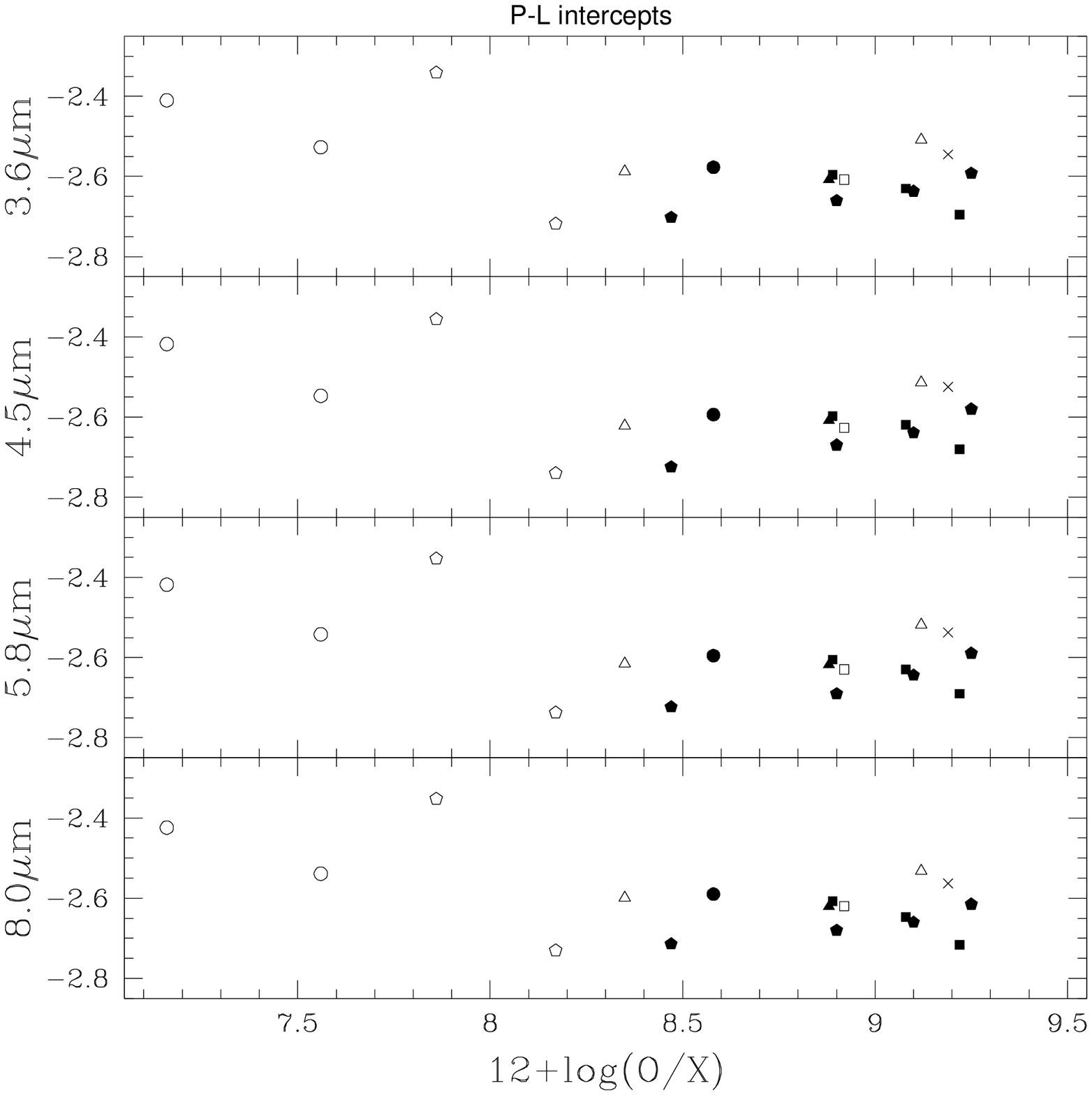}
\caption{Synthetic P-L relations as a function of $12+\log(O/H)$, separated according to the $\Delta Y/\Delta Z$ values. Note that the open symbols are for $\Delta Y/\Delta Z>3$, and filled symbols are for $\Delta Y/\Delta Z$ between $1$ and $3$ (inclusive). \label{pl_zx_delta}}
\end{figure*} 

\begin{figure*}
$\begin{array}{cccc}
\includegraphics[angle=0,scale=0.4]{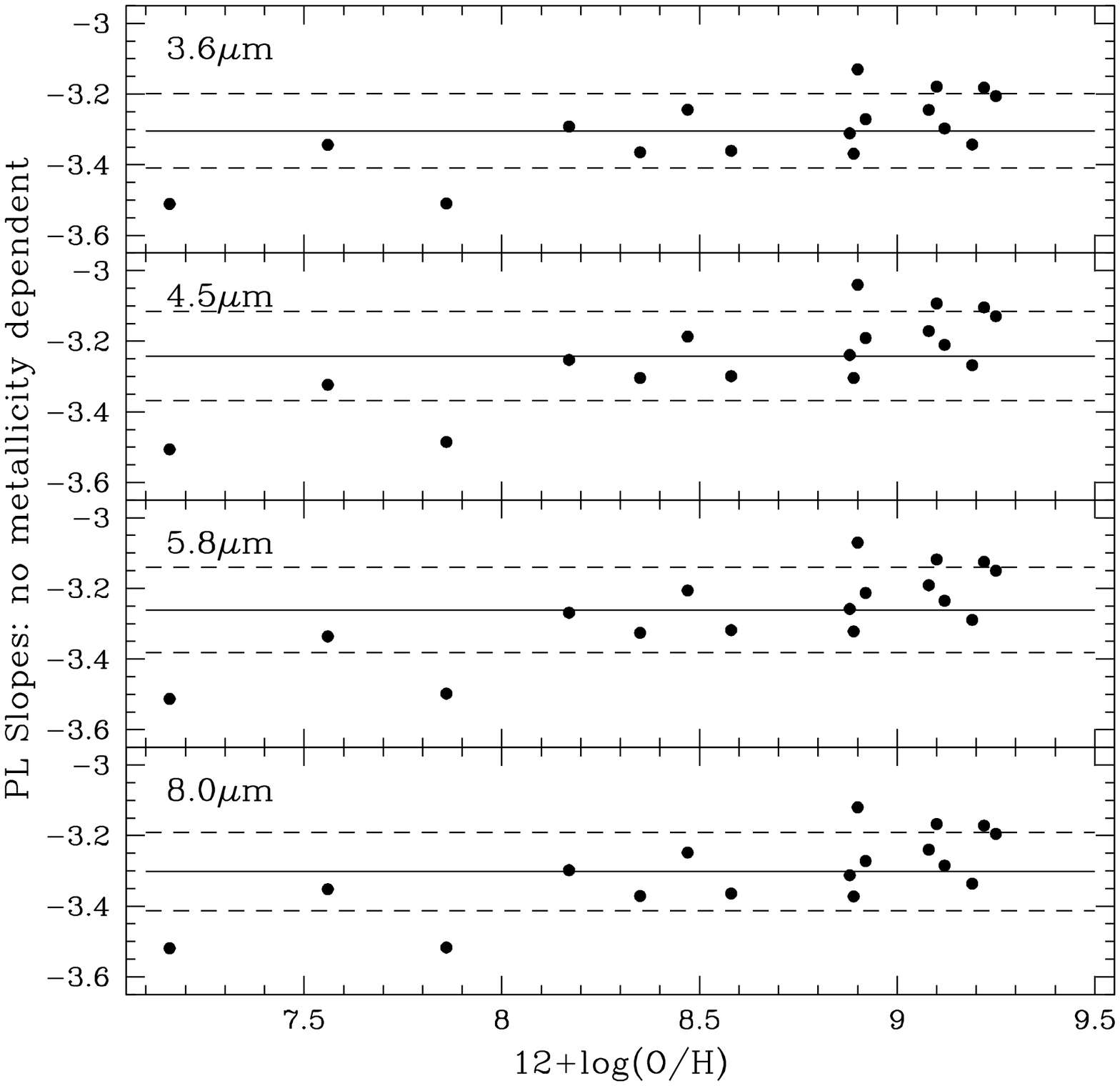} &
\includegraphics[angle=0,scale=0.4]{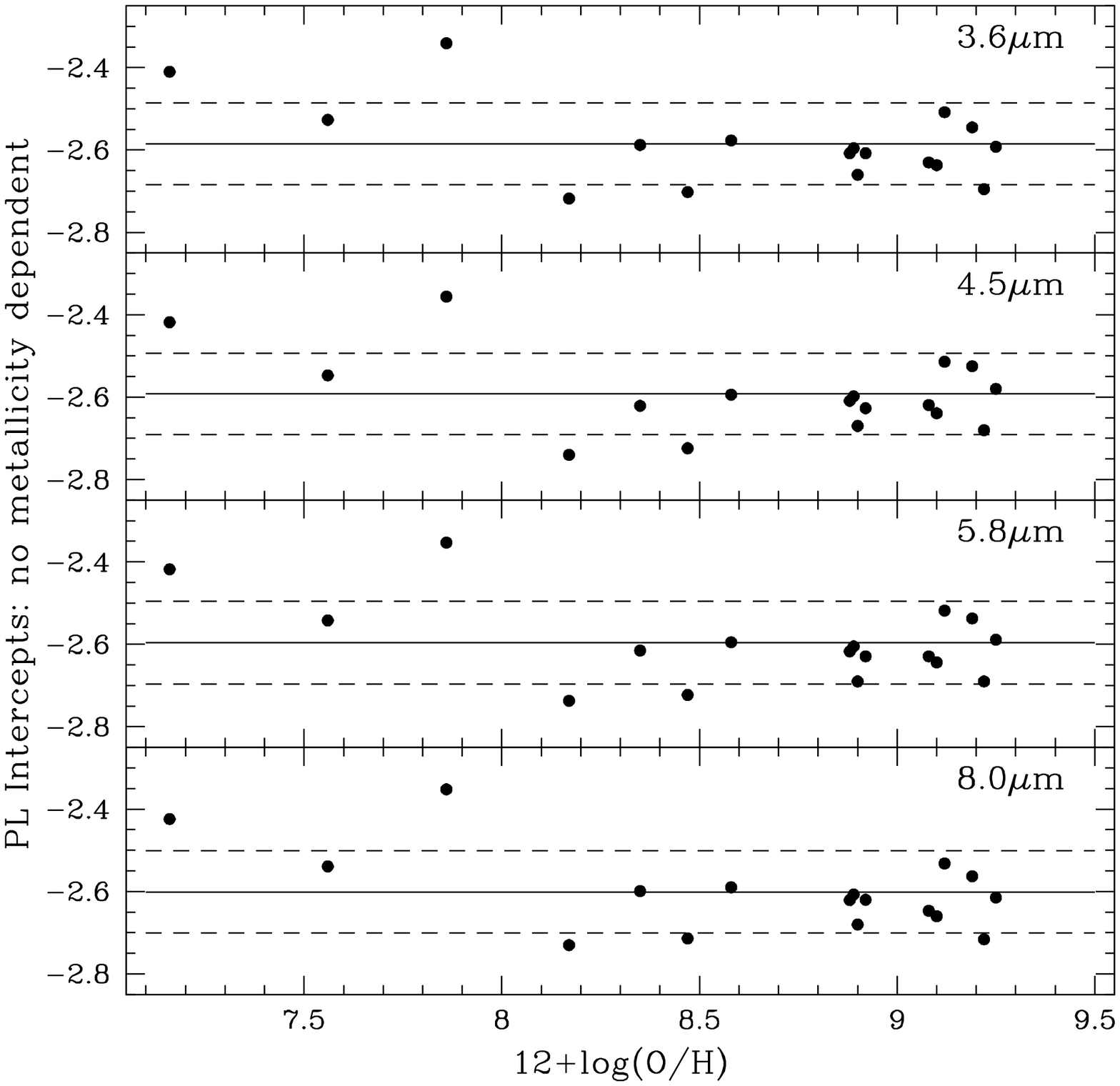} \\
\includegraphics[angle=0,scale=0.4]{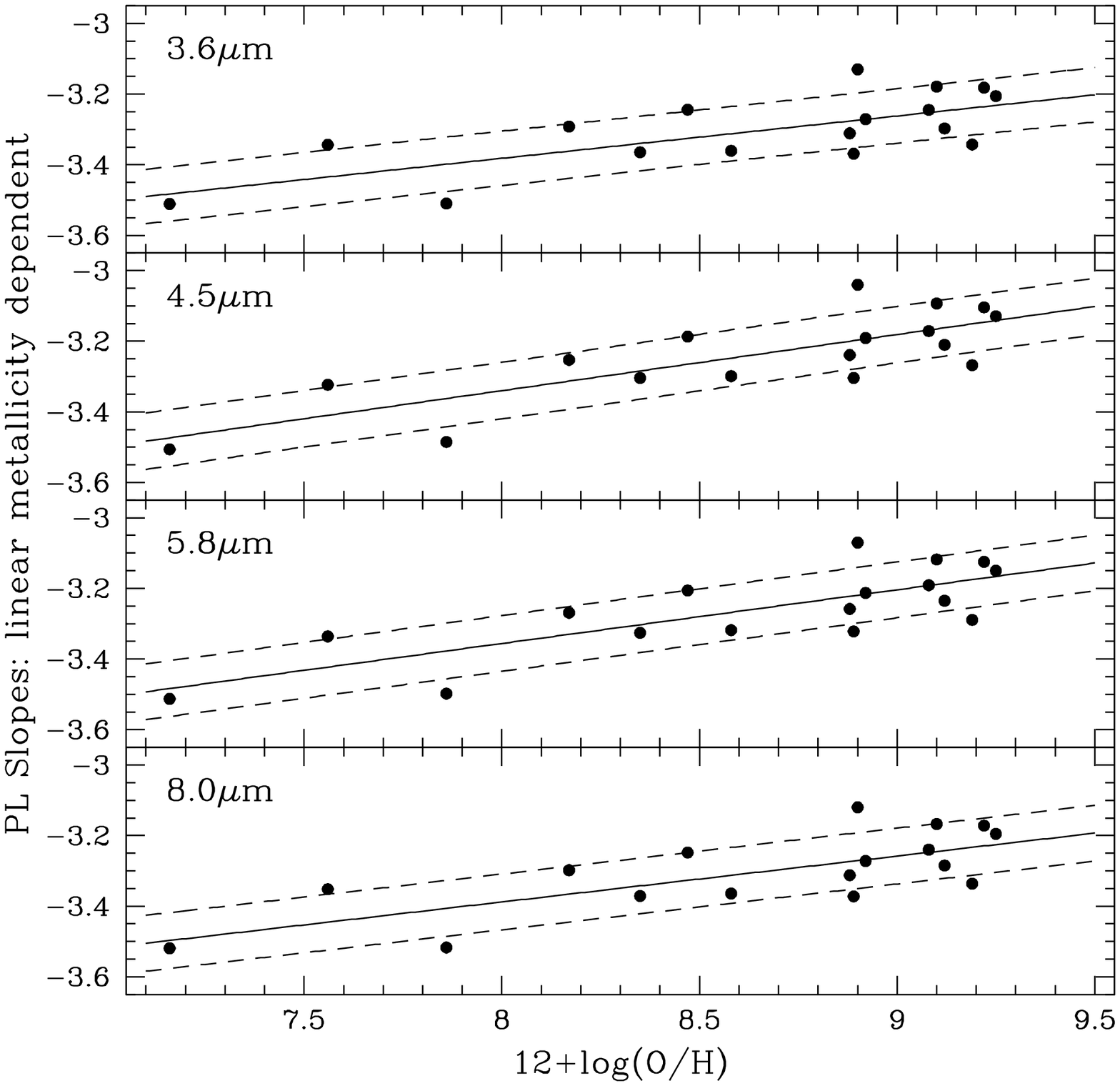} &
\includegraphics[angle=0,scale=0.4]{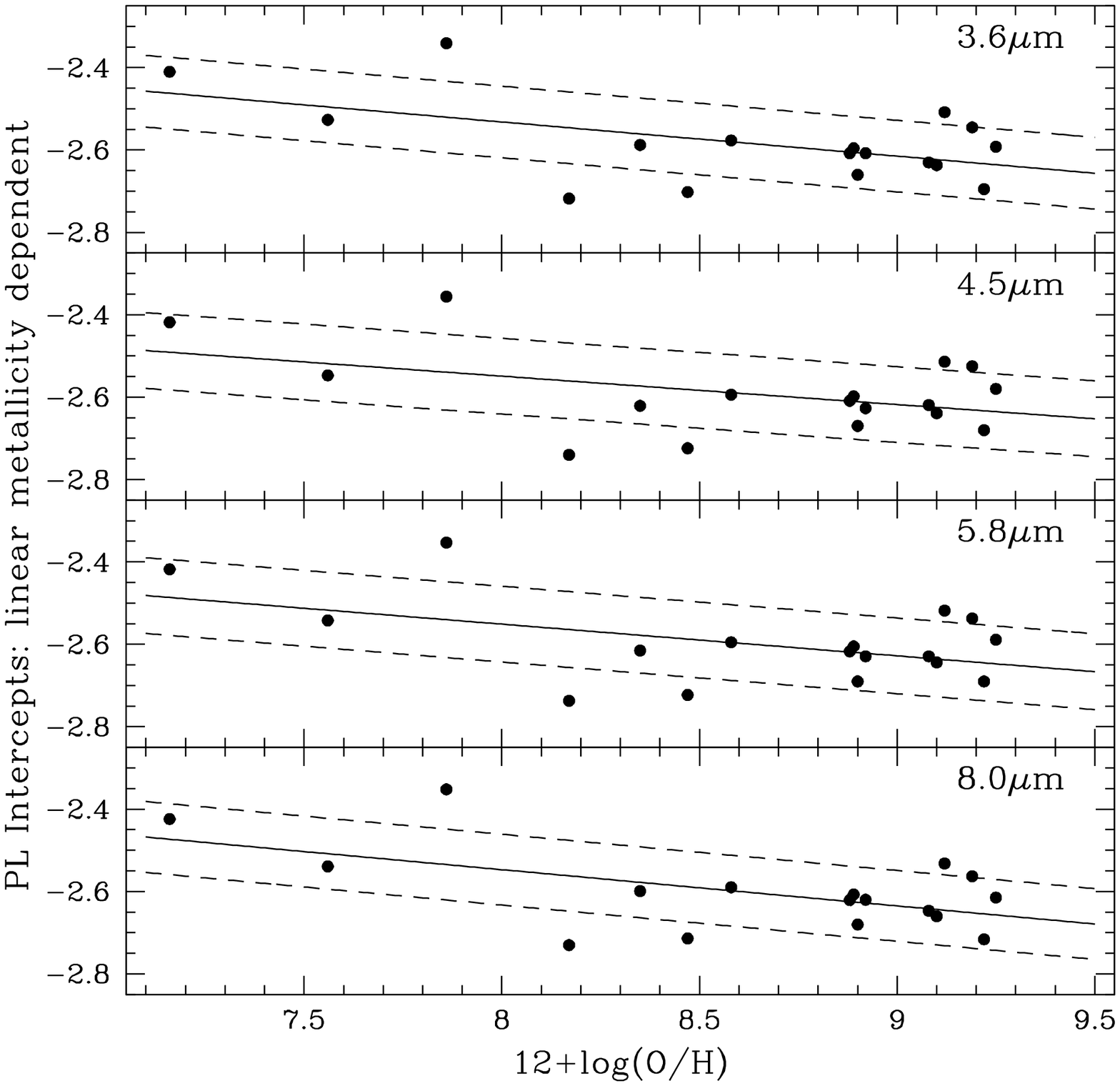} \\
\end{array}$ 
\caption{Fitting of the constant (upper panels) and linear (lower panels) regression models to the synthetic P-L slopes (left panels) and intercepts (right panels).  The dashed lines represent the $1\sigma$ boundary of the fitted regressions. The plots for $[\mathrm{Fe}/H]$ are very similar to this figure. \label{pl_rm_oh}}
\end{figure*} 

As mentioned in the Introduction, the IRAC band P-L relations are expected to be insensitive to metallicity. However, based on the residual analysis from multi-band empirical P-L relations for a number of LMC Cepheids with $[\mathrm{Fe}/H]$ measurements, \citet{fre11} suggested the mid-infrared P-L relation could be mildly depending on metallicity. Therefore, possible metallicity dependence of the IRAC band P-L relations is investigated in this section using the synthetic P-L relations presented in Tables \ref{plslope} and \ref{plzp}. Figure \ref{pl_zx} and \ref{pl_zx_delta} present the synthetic P-L slopes (left panels) and intercepts (right panels) as a function of $12+\log(O/H)$\footnote{Plots for the synthetic P-L relations as a function of $\log(Z/X)$ and $[\mathrm{Fe}/H]$ are similar to these figures, hence omitted from the present paper.}, suggesting the IRAC band P-L relations may be sensitive to metallicity and/or helium abundance of the Cepheids.

Well known statistical tests can be used to test the possible metallicity dependence of the synthetic P-L relations as shown in Figures \ref{pl_zx} and \ref{pl_zx_delta}. We considered a null hypothesis that the IRAC band P-L relations are independent of metallicity, which can be represented by a constant regression model in the form of $A=a_0$, where $A$ stands for either the P-L slopes or intercepts. The alternate hypothesis is that there is a linear dependency of metallicity on P-L relations represented by a linear regression model, $A=a_0+a_1B$ where $B=12+\log(O/H)$ or $B=[\mathrm{Fe}/H]$. Graphic representations of these regression models are presented in Figure \ref{pl_rm_oh}. It is worth to point out that these regression models are used to test the dependency of metallicity on synthetic P-L relations, and do not represent the actual linear dependency of metallicity on P-L relations, since the metallicity effect, if present, will not be a simple linear relation (as evident in Figure \ref{pl_zx} and \ref{pl_zx_delta}). It is well-known in statistical literature \citep[for example, see][]{kut05} that the $F$-test can be applied to test if additional parameter (in our case, the $a_1$) is needed in the regression model. By adopting $\alpha=0.05$, the null hypothesis can be rejected if $F>4.54$. The $F$-test results are summarized in Table \ref{ftest1}, showing that the null hypothesis can be rejected in most cases, except for $4.5\mu{\mathrm m}$ and $5.8\mu{\mathrm m}$ band P-L intercepts. These suggested that the synthetic P-L relations may not be independent of metallicity.

\section{The Synthetic IRAC Band P-C Relations}

\begin{figure}
\plotone{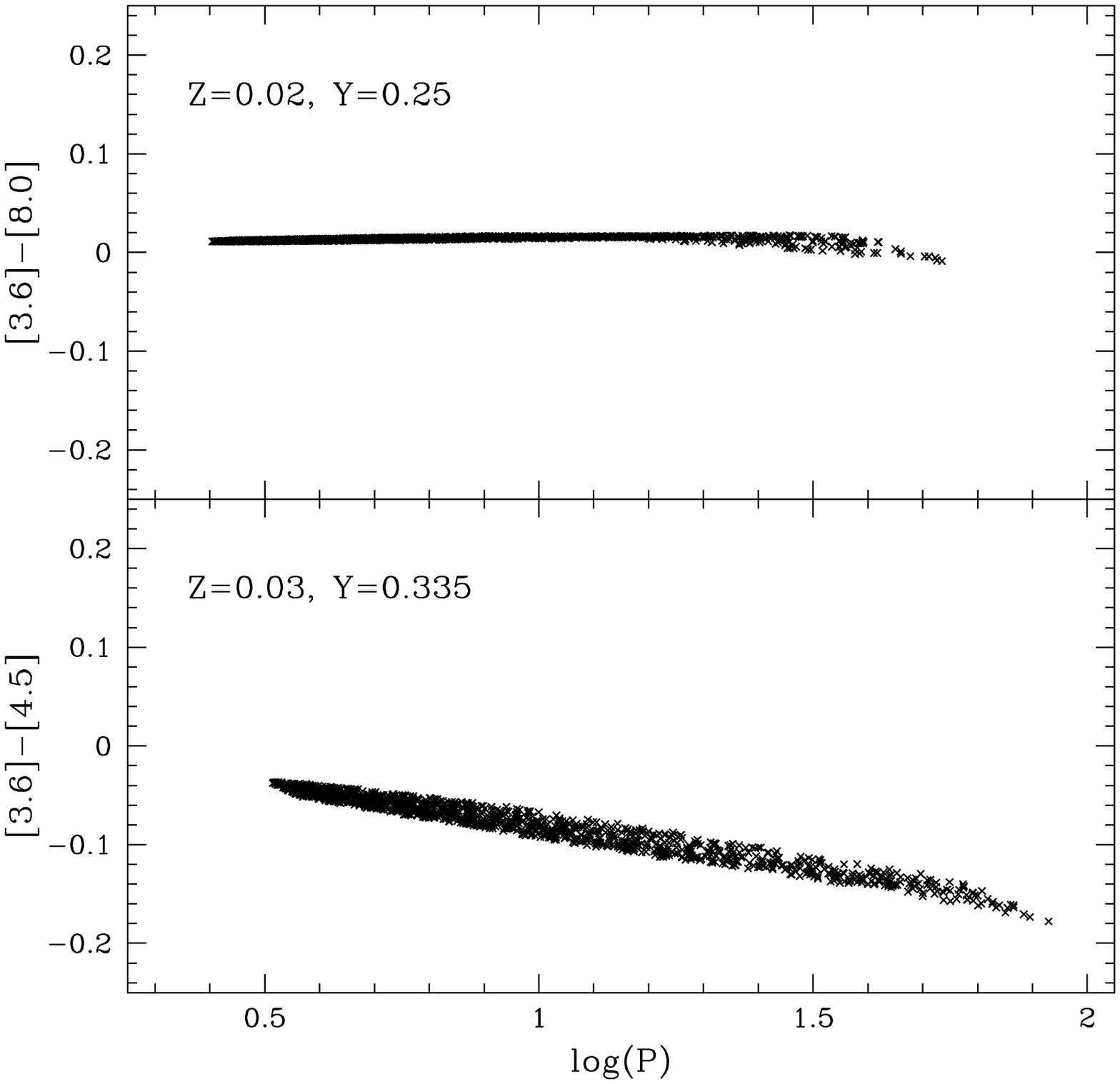}
\caption{Two examples of the synthetic P-C relations in the IRAC band. Top and bottom panels show the example of a flat P-C relation and a P-C relation with largest slope, respectively. The corresponding pulsational model sets are given in the upper-left corners. \label{pc_example}}
\end{figure} 

\begin{figure*}
\plottwo{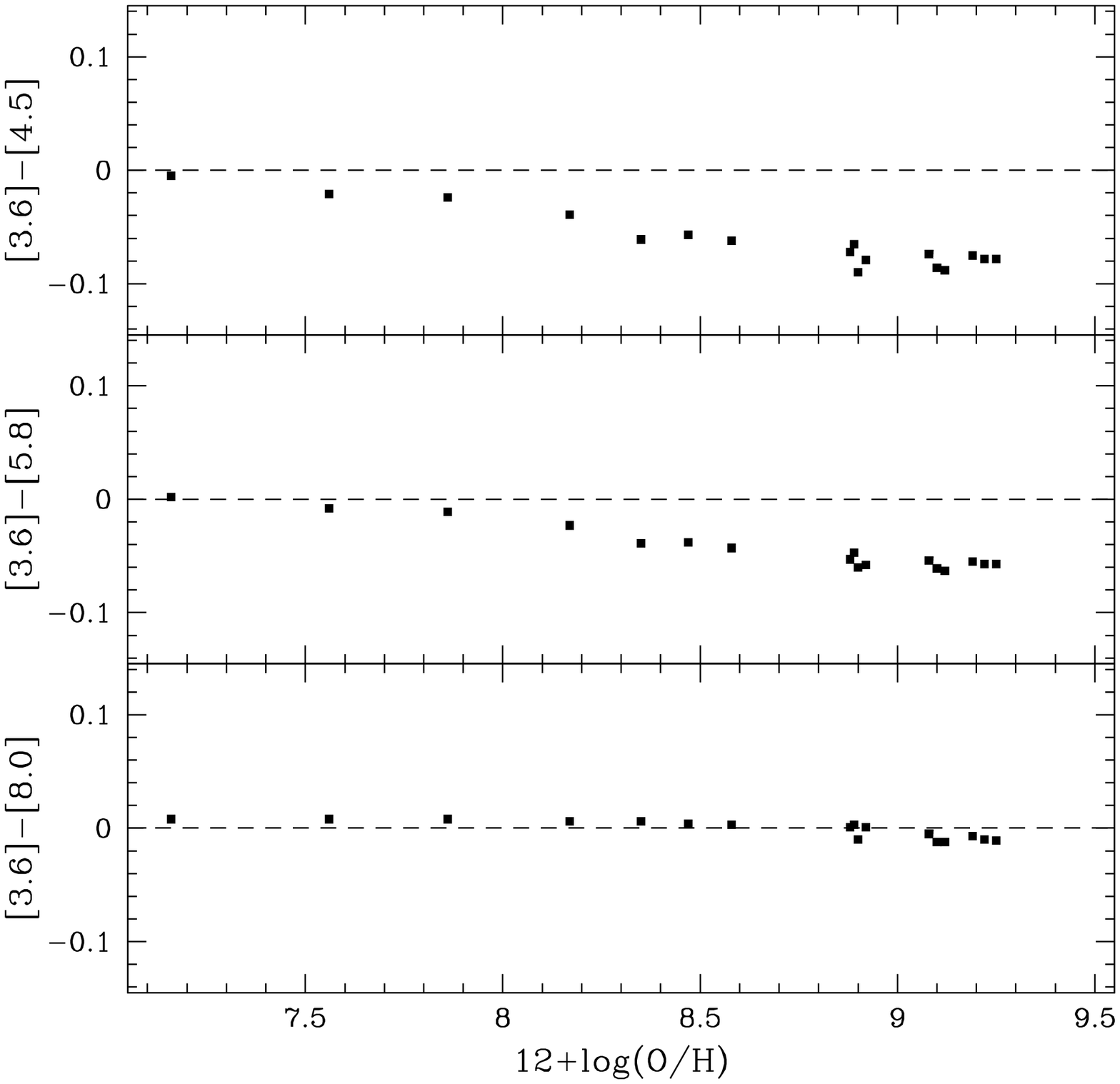}{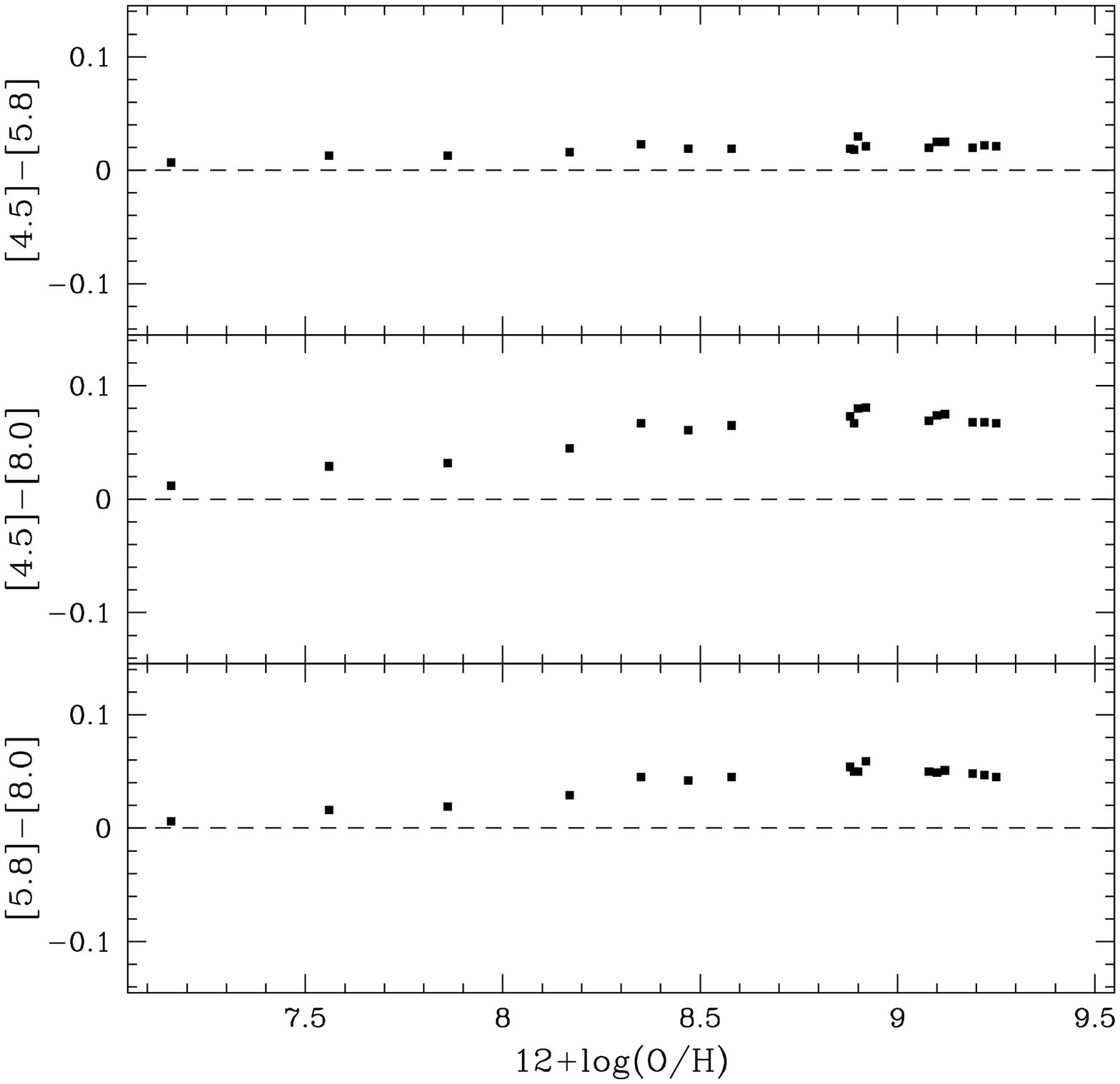}
\caption{Slopes of the synthetic P-C relations as a function of $12+\log(O/H)$. The dashed lines represent $y=0$ in order to guide the eyes, and {\it not} the fitting to the data. \label{pcslope}}
\end{figure*}

\begin{figure*}
\plottwo{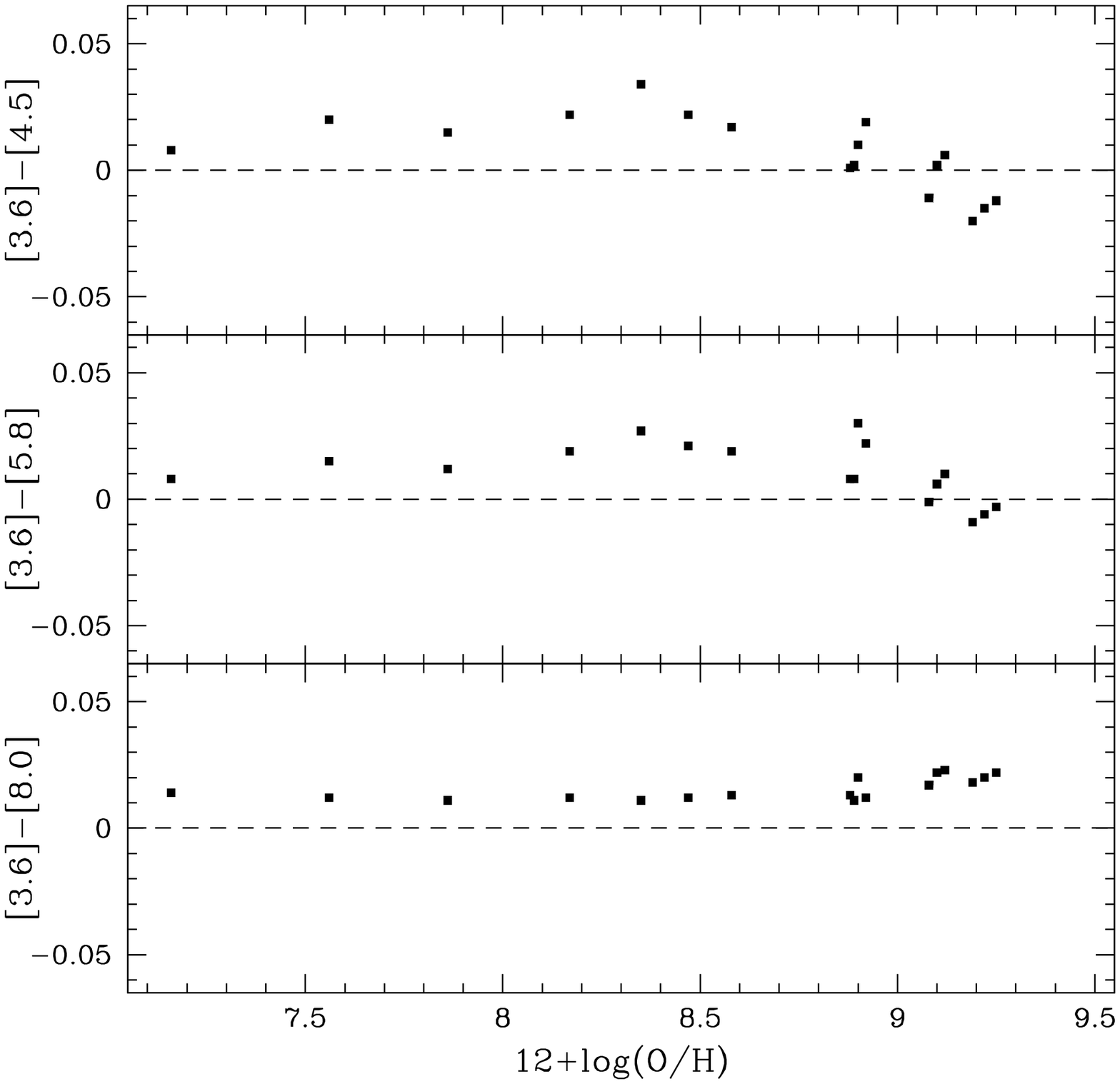}{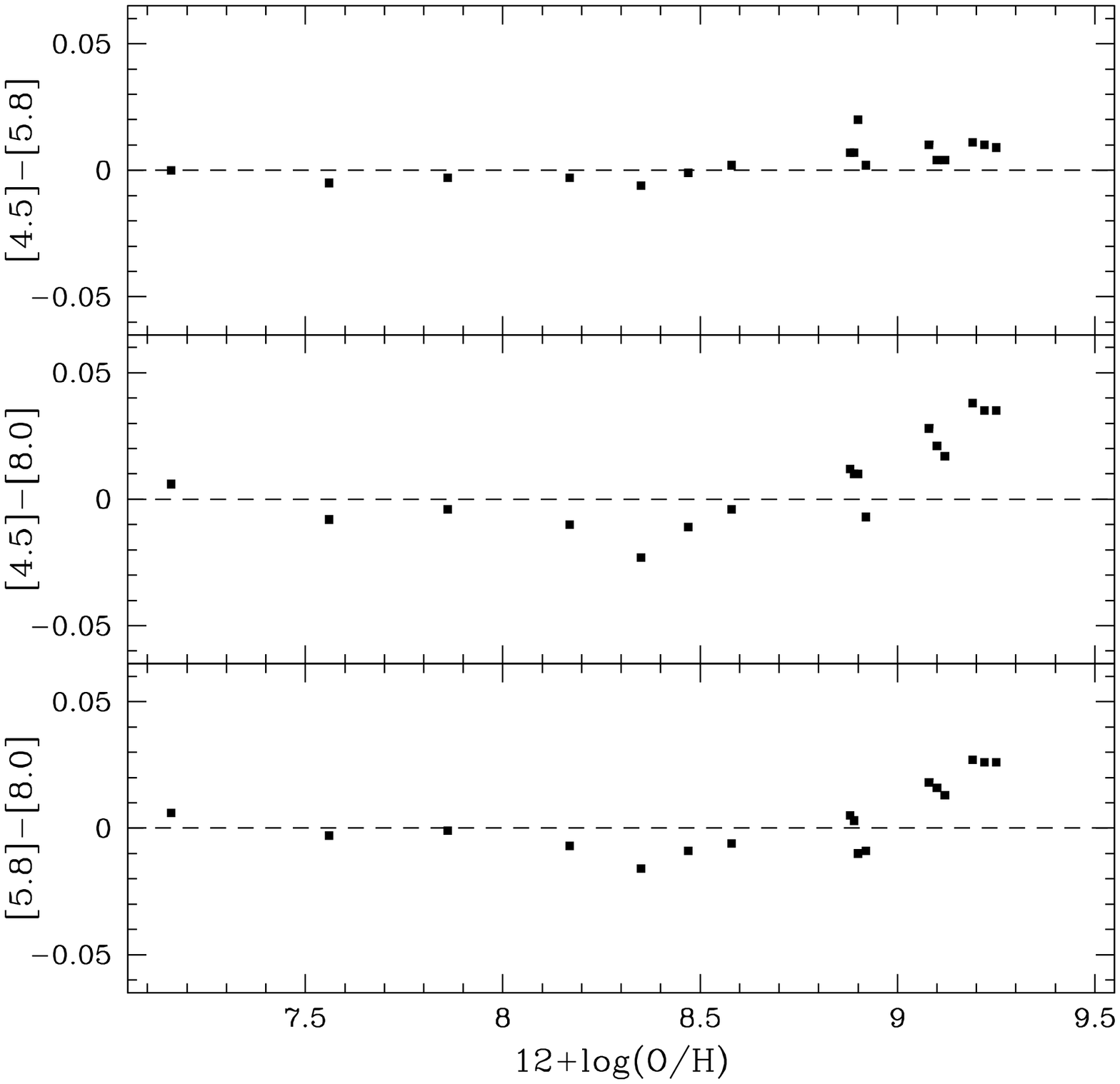}
\caption{Same as Figure \ref{pcslope}, but for the intercepts of the synthetic P-C relations. \label{pczp}}
\end{figure*}

Pulsators from various model sets, as described in Section 2, can also be used to construct the synthetic P-C relations in the IRAC bands. For brevity, colors from $3.6\mu{\mathrm m}$ and $4.5\mu{\mathrm m}$ bands are denoted as $[3.6]-[4.5]$, and so on. Two examples of the synthetic P-C relations are presented in Figure \ref{pc_example}. Results of the synthetic P-C relations are summarized in Table \ref{tab_pc}. As in the case of the synthetic P-L relations, pulsators with  $0.4 \leq \log(P) \leq 2.0$ were used to fit the P-C relations. The P-C slopes and intercepts do not deviate by more than $0.006$ if all the pulsators were included. The synthetic P-C slopes and intercepts were plotted as a function of $12+\log(O/H)$ in Figures \ref{pcslope} and \ref{pczp}, respectively. From these figures, it is clear that P-C relations exist for certain combinations of IRAC band filters and metallicity, and not all of the synthetic P-C relations are independent of metallicity. The P-C relations that are independent or insensitive to metallicity are the $[3.6]-[8.0]$ and $[4.5]-[5.8]$ P-C relations, especially for those with $12+\log(O/H)<8.9$.

\begin{deluxetable*}{llrrrrrr}
\tabletypesize{\scriptsize}
\tablecaption{Synthetic IRAC band P-C relations at various metallicities\tablenotemark{a}. \label{tab_pc}}
\tablewidth{0pt}
\tablehead{
\colhead{$Z$} &
\colhead{$Y$} & 
\colhead{$[3.6]-[4.5]$} &
\colhead{$[3.6]-[5.8]$} &
\colhead{$[3.6]-[8.0]$} &
\colhead{$[4.5]-[5.8]$} &
\colhead{$[4.5]-[8.0]$} &
\colhead{$[5.8]-[8.0]$} 
}
\startdata
\cutinhead{P-C Slopes} 
$0.0004$ & $0.24$ &$-0.005$& $0.002$ & $0.008$ & $0.007$ & $0.012$ & $0.006$ \\
$0.001$  & $0.24$ &$-0.021$&$-0.008$ & $0.008$ & $0.013$ & $0.029$ & $0.016$ \\
$0.002$  & $0.24$ &$-0.024$&$-0.011$ & $0.008$ & $0.013$ & $0.032$ & $0.019$ \\
$0.004$  & $0.25$ &$-0.039$&$-0.023$ & $0.006$ & $0.016$ & $0.045$ & $0.029$ \\   
$0.006$  & $0.25$ &$-0.061$&$-0.039$ & $0.006$ & $0.023$ & $0.067$ & $0.045$ \\
$0.008$  & $0.25$ &$-0.057$&$-0.038$ & $0.004$ & $0.019$ & $0.061$ & $0.042$ \\
$0.01$   & $0.26$ &$-0.062$&$-0.043$ & $0.003$ & $0.019$ & $0.065$ & $0.045$ \\
$0.02$   & $0.25$ &$-0.072$&$-0.053$ & $0.001$ & $0.019$ & $0.073$ & $0.054$ \\
$0.02$   & $0.26$ &$-0.065$&$-0.047$ & $0.003$ & $0.018$ & $0.067$ & $0.050$ \\
$0.02$   & $0.28$ &$-0.09$\tablenotemark{b} &$-0.06$\tablenotemark{b}  & $-0.01$\tablenotemark{b} & $0.03$\tablenotemark{b}  & $0.08$\tablenotemark{b}  & $0.05$\tablenotemark{b} \\
$0.02$   & $0.31$ &$-0.079$&$-0.058$ & $0.001$ & $0.021$ & $0.081$ & $0.059$ \\
$0.03$   &$0.275$ &$-0.074$&$-0.054$ &$-0.005$ & $0.020$ & $0.069$ & $0.050$ \\
$0.03$   & $0.31$ &$-0.086$&$-0.061$ &$-0.012$ & $0.025$ & $0.074$ & $0.049$ \\
$0.03$   &$0.335$ &$-0.088$&$-0.063$ &$-0.012$ & $0.025$ & $0.075$ & $0.051$ \\
$0.04$   & $0.25$ &$-0.075$&$-0.055$ &$-0.007$ & $0.020$ & $0.068$ & $0.048$ \\
$0.04$   & $0.29$ &$-0.078$&$-0.057$ &$-0.010$ & $0.022$ & $0.068$ & $0.047$ \\
$0.04$   & $0.33$ &$-0.078$&$-0.057$ &$-0.011$ & $0.021$ & $0.067$ & $0.045$ \\
\cutinhead{P-C Intercepts} 
$0.0004$ & $0.24$ & $0.008$& $0.008$ & $0.014$ &$-0.000$ & $0.006$ & $0.006$ \\
$0.001$  & $0.24$ & $0.020$& $0.015$ & $0.012$ &$-0.005$ &$-0.008$ &$-0.003$ \\
$0.002$  & $0.24$ & $0.015$& $0.012$ & $0.011$ &$-0.003$ &$-0.004$ &$-0.001$ \\
$0.004$  & $0.25$ & $0.022$& $0.019$ & $0.012$ &$-0.003$ &$-0.010$ &$-0.007$ \\
$0.006$  & $0.25$ & $0.034$& $0.027$ & $0.011$ &$-0.006$ & $-0.023$&$-0.016$ \\
$0.008$  & $0.25$ & $0.022$& $0.021$ & $0.012$ &$-0.001$ &$-0.011$ &$-0.009$ \\ 
$0.01$   & $0.26$ & $0.017$& $0.019$ & $0.013$ & $0.002$ &$-0.004$ &$-0.006$ \\
$0.02$   & $0.25$ & $0.001$& $0.008$ & $0.013$ & $0.007$ & $0.012$ & $0.005$ \\
$0.02$   & $0.26$ & $0.002$& $0.008$ & $0.011$ & $0.007$ & $0.010$ & $0.003$ \\
$0.02$   & $0.28$ & $0.01$\tablenotemark{b} & $0.03$\tablenotemark{b}  & $0.02$\tablenotemark{b}  & $0.02$\tablenotemark{b}  & $0.01$\tablenotemark{b}  & $-0.01$\tablenotemark{b} \\ 
$0.02$   & $0.31$ & $0.019$& $0.022$ & $0.012$ & $0.002$ &$-0.007$ &$-0.009$ \\
$0.03$   &$0.275$ &$-0.011$&$-0.001$ & $0.017$ & $0.010$ & $0.028$ & $0.018$ \\
$0.03$   & $0.31$ & $0.002$& $0.006$ & $0.022$ & $0.004$ & $0.021$ & $0.016$ \\
$0.03$   &$0.335$ & $0.006$& $0.010$ & $0.023$ & $0.004$ & $0.017$ & $0.013$ \\
$0.04$   & $0.25$ &$-0.020$&$-0.009$ & $0.018$ & $0.011$ & $0.038$ & $0.027$ \\
$0.04$   & $0.29$ &$-0.015$&$-0.006$ & $0.020$ & $0.010$ & $0.035$ & $0.026$ \\
$0.04$   & $0.33$ &$-0.012$&$-0.003$ & $0.022$ & $0.009$ & $0.035$ & $0.026$  
\enddata
\tablenotetext{a}{Errors of the P-C slopes and intercepts are ignored as they are generally less than $0.003$.}
\tablenotetext{b}{Derived from the P-L relations.}
\end{deluxetable*}

\subsection{Comparison to Empirical P-C Relations}

\begin{figure*}
$\begin{array}{ccc}
\includegraphics[angle=0,scale=0.28]{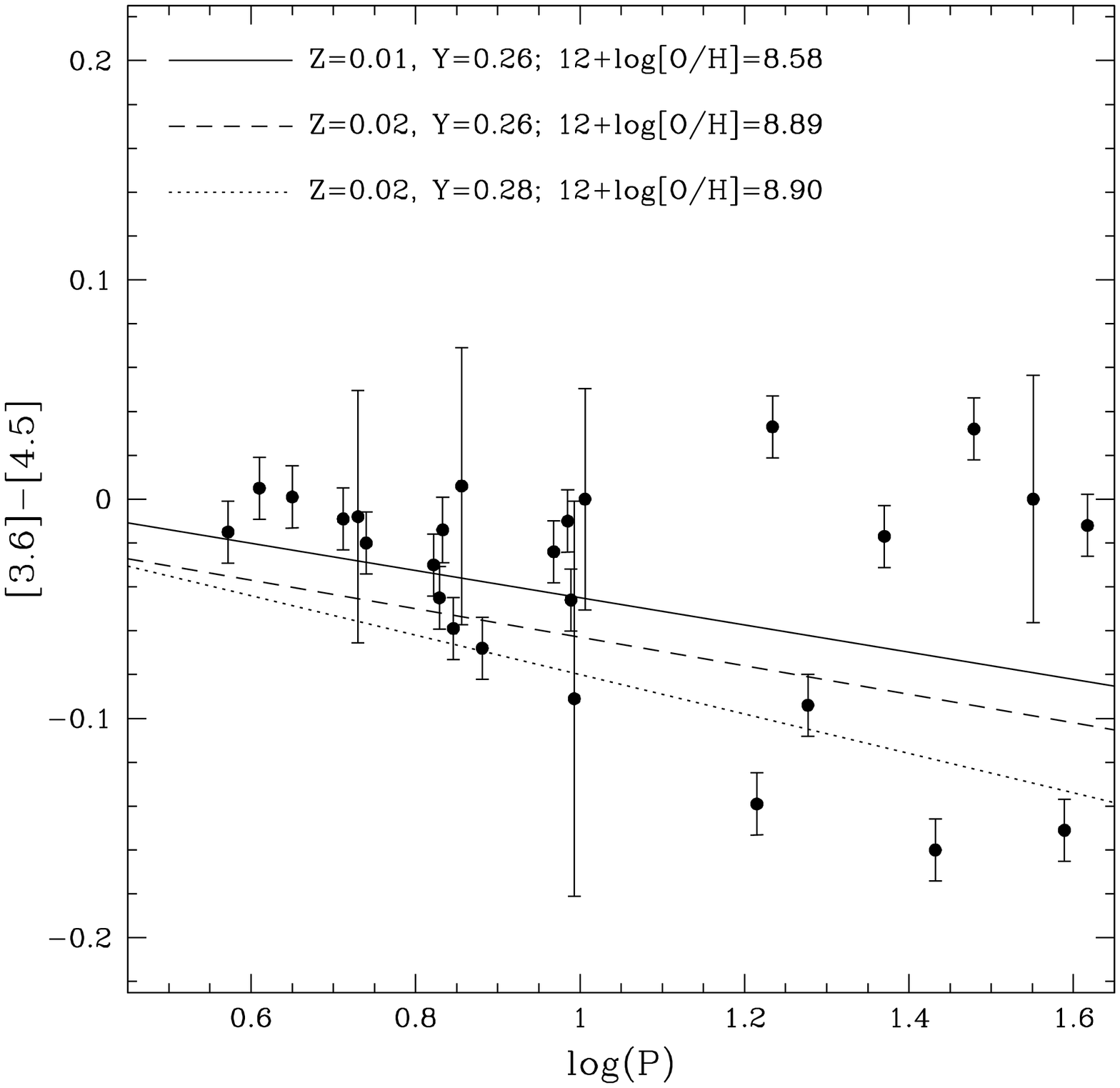} &
\includegraphics[angle=0,scale=0.28]{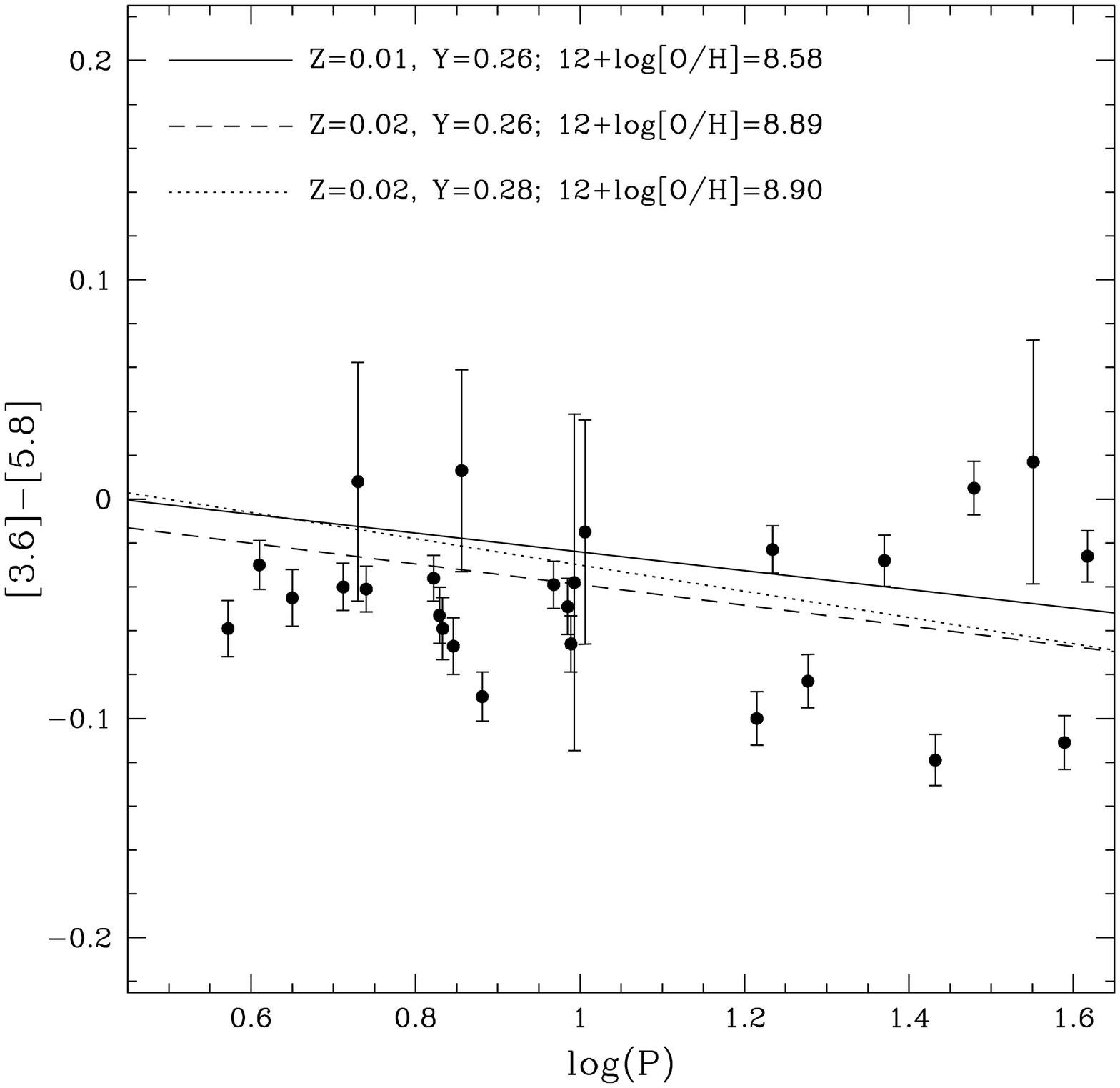} &
\includegraphics[angle=0,scale=0.28]{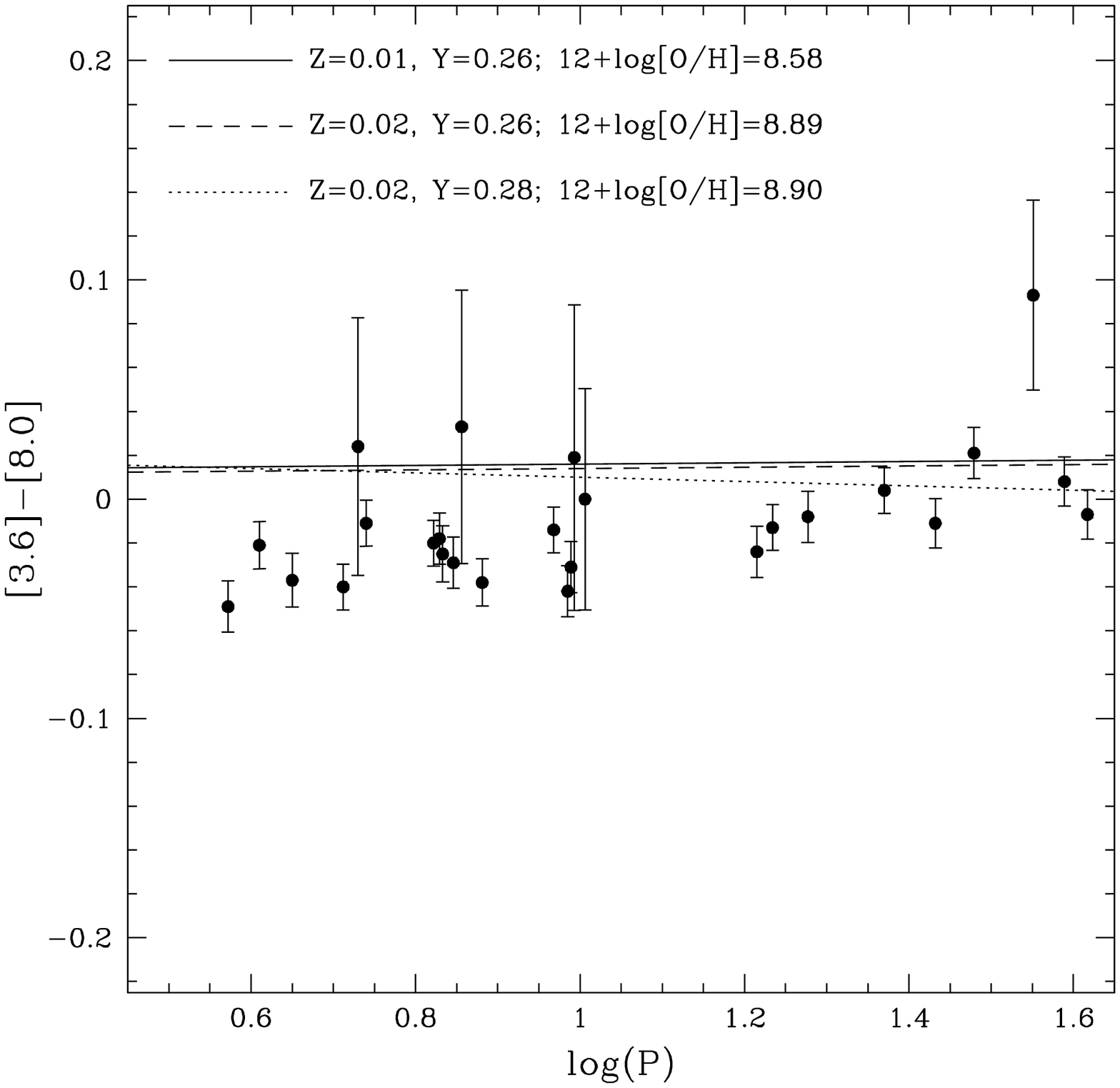} \\
\includegraphics[angle=0,scale=0.28]{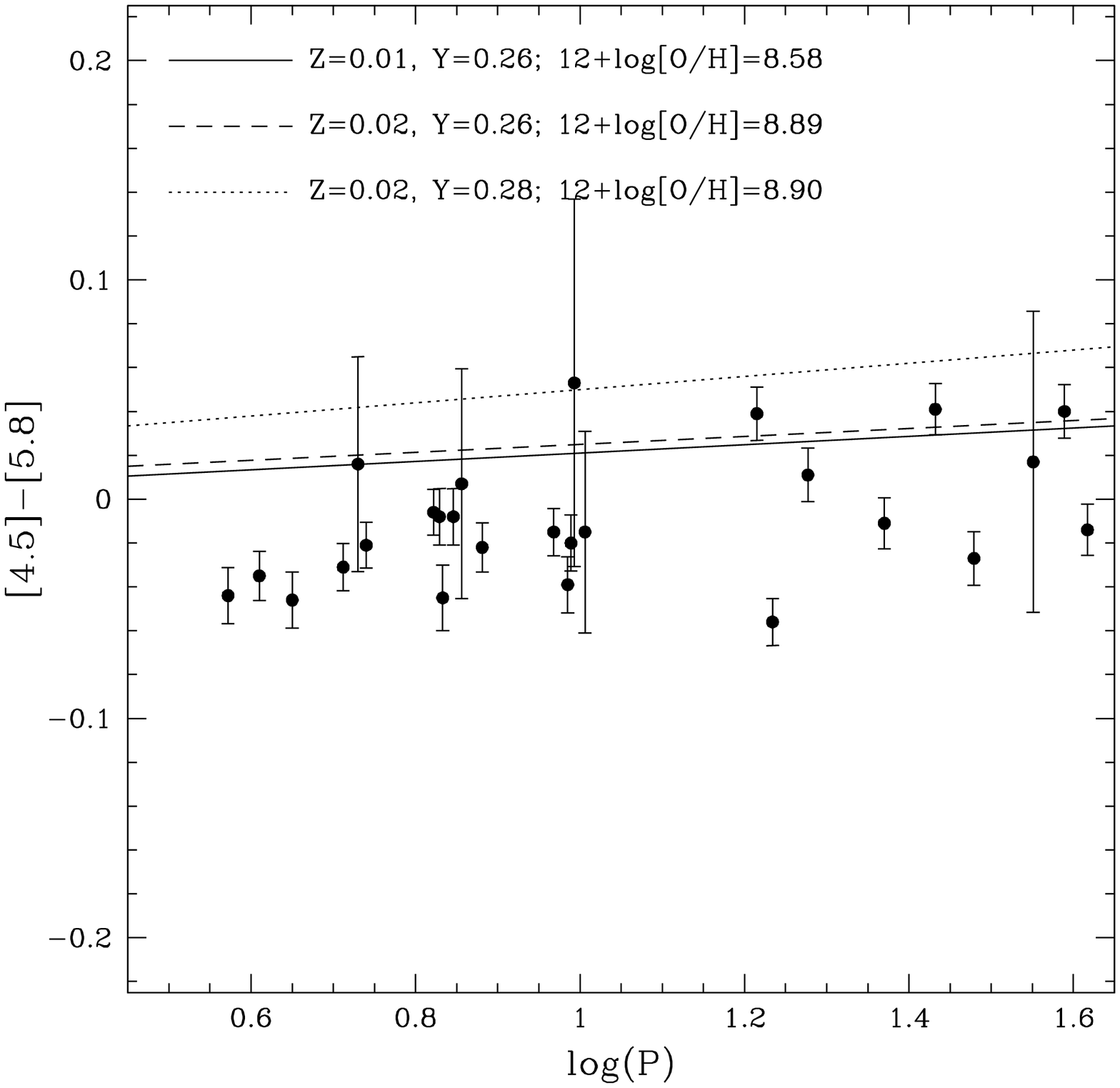} &
\includegraphics[angle=0,scale=0.28]{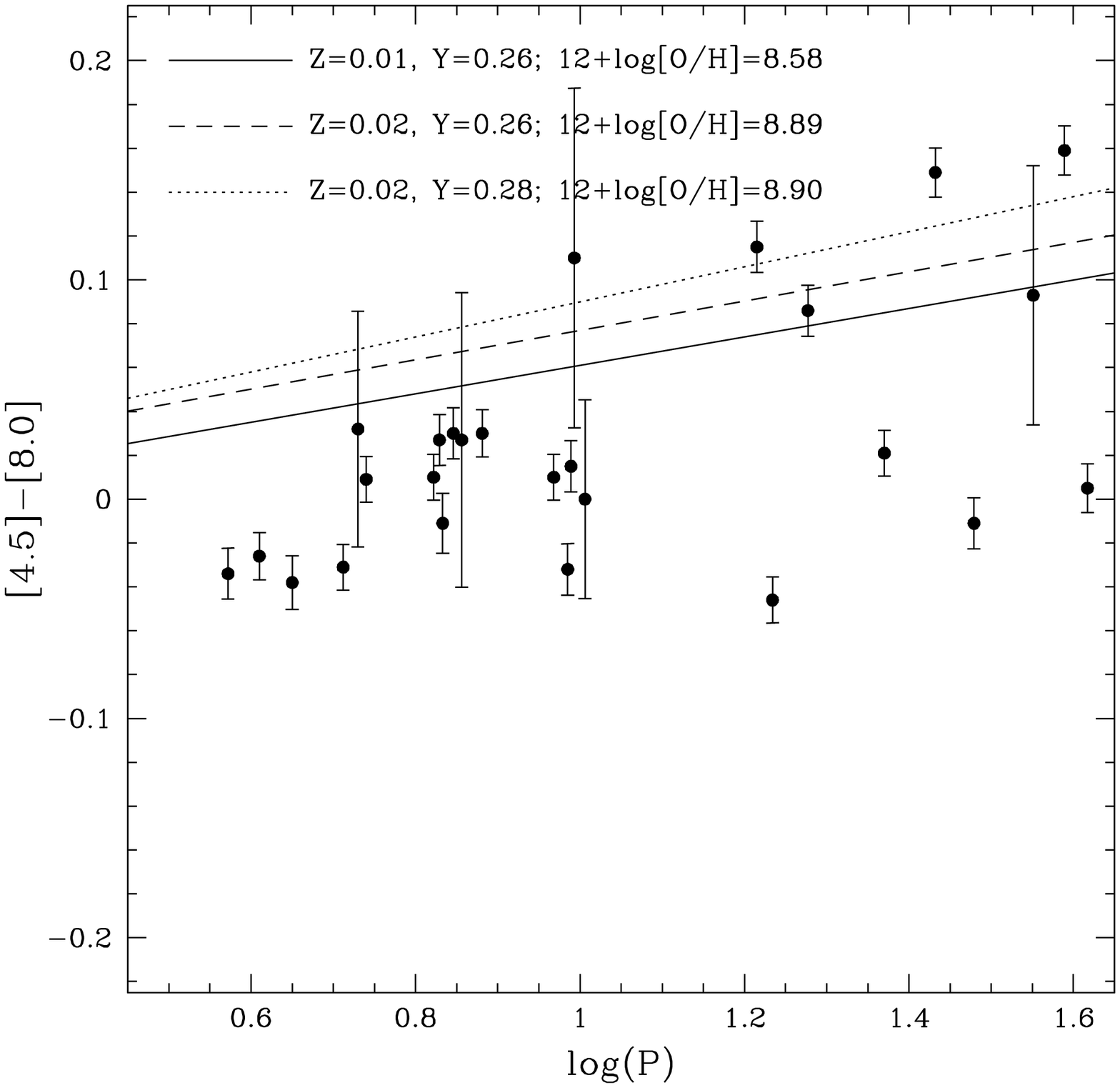} &
\includegraphics[angle=0,scale=0.28]{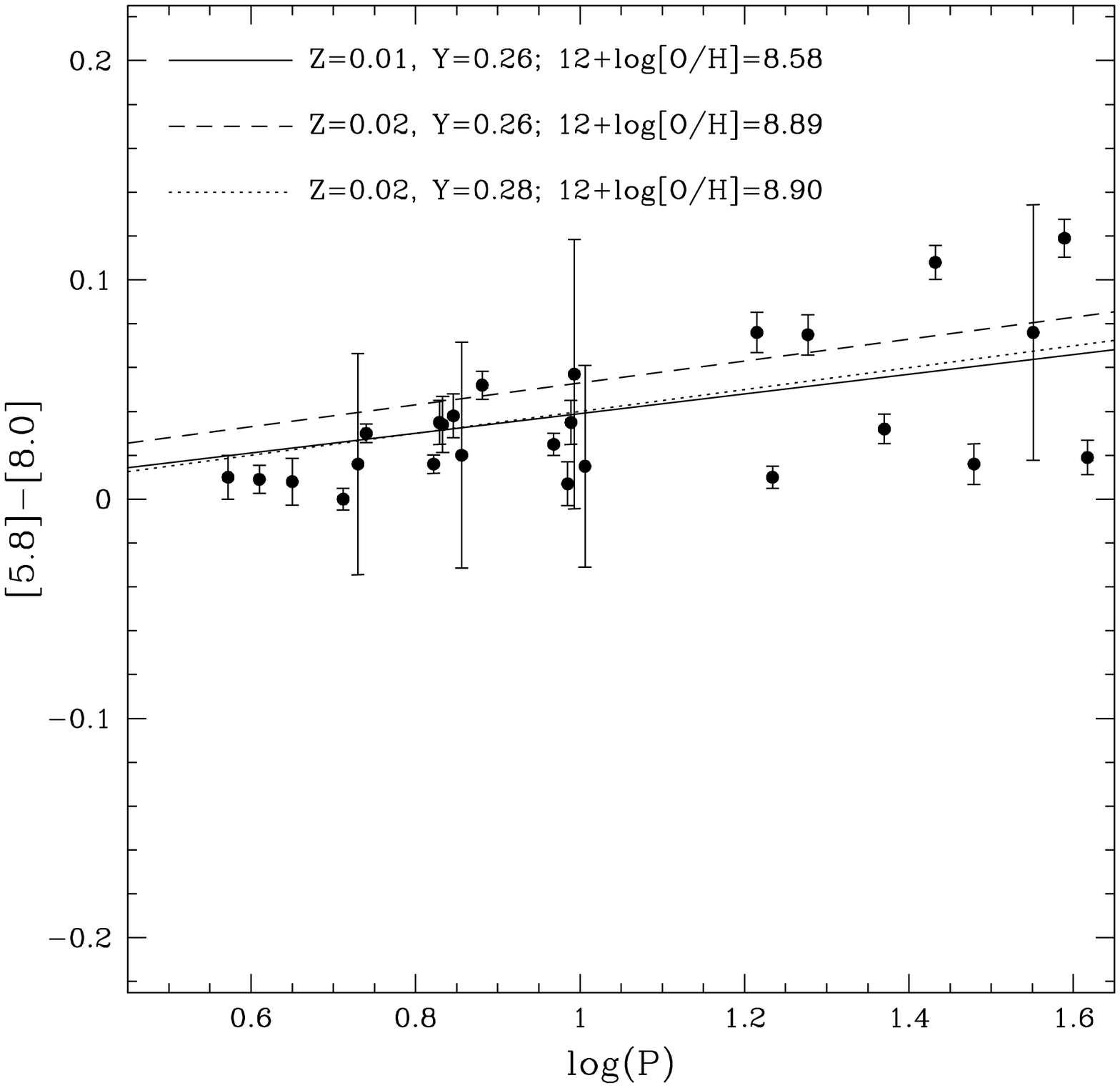} \\
\end{array}$ 
\caption{Comparison of the P-C relations for Galactic Cepheids and the synthetic P-C relations from three selected model sets given in Table \ref{tab_pc}. Photometry for Galactic Cepheids is adopted from \citet{mar10}. \label{gal_pc_compare}}
\end{figure*}

\citet{mar10} found a significant $[3.6]-[8.0]$ P-C relation for the Galactic Cepheids, with the expression of $[3.6]-[8.0]=0.039(\pm0.008)\log(P) - 0.058(\pm0.014)$. This empirical non-zero P-C slope is in contradiction with the synthetic P-C slopes given in Table \ref{tab_pc}, at which the synthetic $[3.6]-[8.0]$ P-C slopes are close to zero. Disagreement was also found for the P-C intercepts. In Figure \ref{gal_pc_compare}, P-C relations for 26 fundamental model Galactic Cepheids were compared to the synthetic P-C relations from selected model sets. IRAC band photometry for these Galactic Cepheids were adopted from \citet{mar10}. This figure shows that only the $[3.6]-[4.5]$ and $[5.8]-[8.0]$ synthetic P-C relations barely agreed with the observed P-C relations. Disagreements between empirical and synthetic P-C relations may due to the small number of Cepheids in the sample, large photometric errors, colors for all Cepheids are not be measured at mean light, presence of circumstellar envelops especially for longer period Cepheids, uncertainties in the adopted model atmospheres or any combinations of these.

\begin{deluxetable}{cccc}
\tabletypesize{\scriptsize}
\tablecaption{Synthetic $[3.6]-[4.5]$ P-C relation for three models sets with pulsators of $1.0<\log (P)<1.8$. \label{tab_pclmc}}
\tablewidth{0pt}
\tablehead{
\colhead{$Z$} &
\colhead{$Y$} &
\colhead{P-C Slope} & 
\colhead{P-C Intercept}  
}
\startdata
0.004 & 0.25 & $-0.072\pm0.003$ & $0.066\pm0.004$ \\
0.006 & 0.25 & $-0.098\pm0.004$ & $0.082\pm0.005$ \\
0.008 & 0.25 & $-0.088\pm0.004$ & $0.065\pm0.005$ 
\enddata
\end{deluxetable}

\begin{figure*}
\plottwo{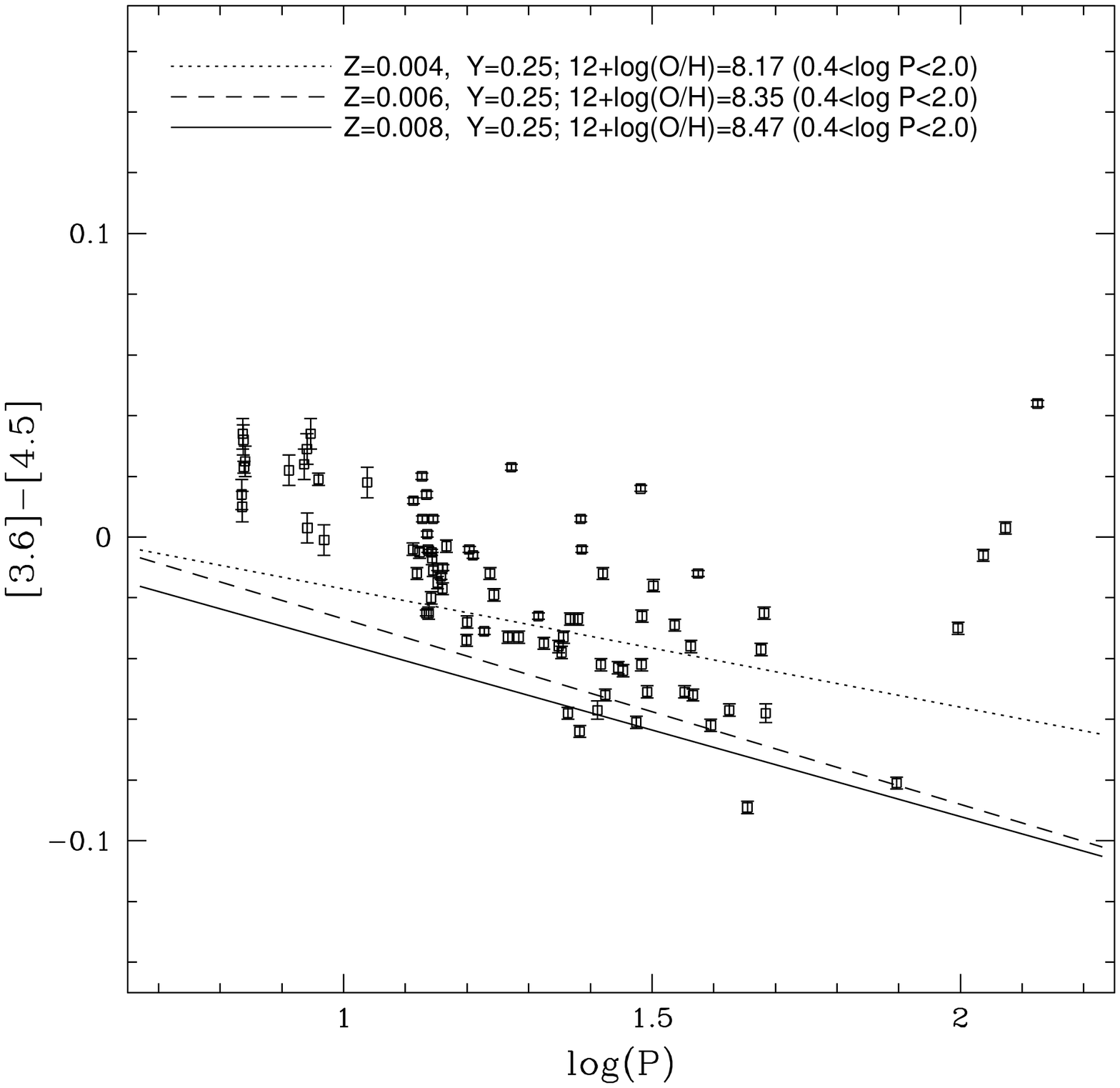}{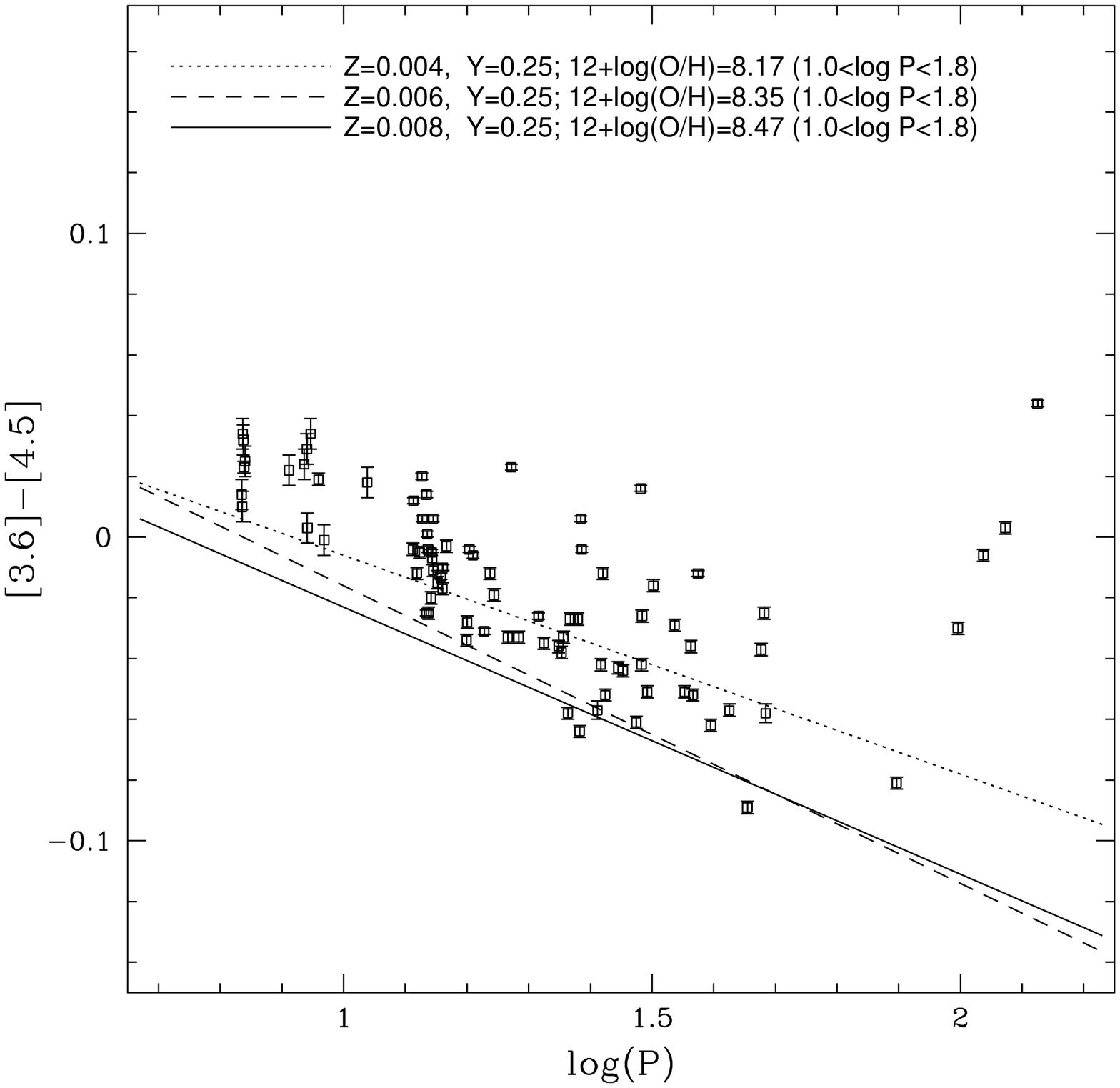}
\caption{Comparison of the P-C relation for LMC Cepheids based on the data presented in \citet{sco11} and the synthetic P-C relations from three selected model sets. Left and right panels show the synthetic P-C relations using pulsators in the period range of $0.4<\log (P)<2.0$ and $1.0<\log (P)<1.8$, respectively. \label{pc_lmc}}
\end{figure*} 

In addition to Galactic Cepheids, \citet{sco11} also reported the finding of $[3.6]-[4.5]$ P-C relation based on LMC Cepheids, within the period range of $1.0<\log (P)<1.8$, that possess multi-epoch observations. The feft panel of Figure \ref{pc_lmc} compares the P-C relation for these Cepheids and the three synthetic P-C relations given in Table \ref{tab_pc}, where the $[3.6]-[4.5]$ colors are obtained from the $3.6\mu{\mathrm m}$ and $4.5\mu{\mathrm m}$ band mean magnitudes taken from \citet{sco11}. It is clear that these synthetic P-C relations do not agree with the empirical P-C relation of $[3.6]-[4.5]=-0.087(\pm0.012)\log(P)+0.092(\pm0.013)$ from \citet{sco11}. However, as in Section 4.2, this empirical P-C relation should also be compared to the synthetic P-C relations based on pulsators within the period range of $1.0<\log (P)<1.8$. These synthetic P-C relations for the three model sets are summarized in Table \ref{tab_pclmc}, and compared to the empirical P-C relation in the right panel of Figure \ref{pc_lmc}. Table \ref{tab_pclmc} shows that the synthetic P-C slope from ($Z=0.008$, $Y=0.25$) model set is in excellent agreement to the empirical P-C slope, although the synthetic P-C intercept is $\sim 2\sigma$ smaller than the empirical counterpart. On the other hand, synthetic P-C relation from ($Z=0.006$, $Y=0.25$) model set is also consistent with the empirical P-C relation.

\subsection{The Wesenheit Function}

Existence of the P-C relations suggested the Wesenheit function can be formulated in the IRAC bands. In optical bands, dispersion of the Wesenheit function is $\sim2$ to $\sim3$ times smaller than the optical P-L relations \citep[for example, see][and reference therein]{fou07,sos08,nge09}. The dispersions of the synthetic Wesenheit function, in the form of $W=m_{3.6\mu{\mathrm m}}-3.40\times ([3.6]-[4.5])$, from all model sets are on average about $\sim1.5$ larger than the dispersions from the synthetic $3.6\mu{\mathrm m}$ band P-L relations. Also, it is not worth to derive the IRAC band Wesenheit function because Wesenheit function is reddening-free by definition. On the other hand, extinction can almost be ignored for the IRAC band P-L relations \citep{fre08,fre10,nge09,mar10,fre11b}. Therefore, there is no net gain by using the Wesenheit function in the IRAC bands.

\section{Discussion and Conclusion}

Synthetic IRAC band P-L relations based on a series of pulsation models were investigated in this paper. These synthetic P-L relations were compared to their empirical counterparts, and the possible metallicity dependency of the IRAC band P-L relations was examined. For the former part, selected sets of synthetic IRAC band P-L relations show agreement to the empirical P-L relations derived from Galactic and Magellanic Cloud Cepheids. The $BVIJK$ synthetic P-L relations for these selected model sets also agreed with their optical and near-infrared counterparts, as presented in \citet{bon10}, for the Galactic and Magellanic Clouds P-L relations. For the metallicity dependency of IRAC band P-L relations, plots for all of the synthetic P-L relations as a function of metallicity revealed that the IRAC band P-L relations may not be independent of metallicity. This result is also supported by statistical $F$-test. On the other hand, current empirical IRAC band P-L relations based on three galaxies, either the P-L relations in ``steep'' or ``shallow'' groups \citep[see Table 2 of][]{nge10}, suggested the IRAC band P-L relations may not be depending on metallicity, or have a weak dependency on metallicity. It is clear that more empirical determinations of the IRAC band P-L relations are needed, especially for galaxies with low metallicity, from future observations with {\it JWST}. 

\begin{figure}
\plotone{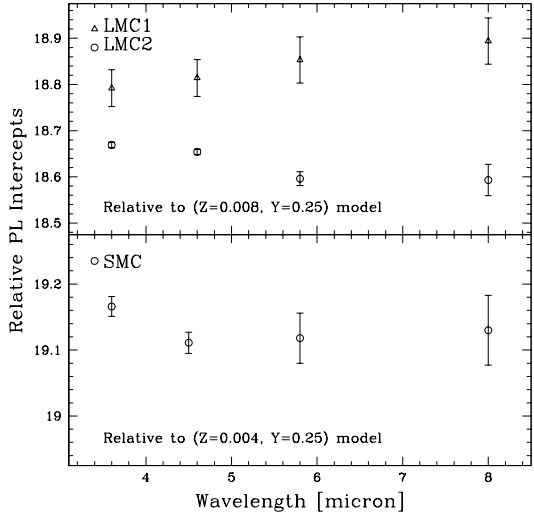}
\caption{Comparison of the P-L intercepts between the empirical P-L relations and the synthetic P-L relations from selected model sets. The top panel shows the comparison for LMC, with synthetic P-L relations selected from ($Z=0.008$, $Y=0.25$) model set. The bottom panel shows the comparison for SMC, with synthetic P-L relations selected from ($Z=0.004$, $Y=0.25$) model set. \label{relzp}}
\end{figure}

Figure \ref{relzp} shows the intercepts for LMC and SMC empirical P-L relations relative to the two adopted synthetic P-L relations, which is equivalent to derive the distance moduli to the Magellanic Clouds using the synthetic P-L relations. The resulted distance moduli for LMC and SMC are higher than the values commonly adopted in literature (e.g., $18.5$ mag and $19.0$ mag for LMC and SMC, respectively). This is due to the already mentioned limitation of the adoption of a canonical M-L relation. However, previous theoretical computations of current nonlinear convective models \citep[see, e.g.,][and references therein]{bon02,cap02,bon08} have shown that by relying on non-canonical models based on an M-L relation brighter than the canonical one by $0.25$ dex the inferred distance moduli are shorter than the values obtained in the canonical scenario by $0.15$-$0.2$ mag depending on the filters. This implies that adopting non-canonical relations in the comparison with empirical {\it Spitzer} data we would have obtained shorter distance moduli by $0.15$-$0.2$ mag for each selected chemical composition, in better agreement with most recent and adopted values in the literature.

Finally, synthetic P-C relations in the IRAC band were also derived and compared to the Galactic and LMC empirical counterparts. In general, disagreements were found between the synthetic and empirical P-C relations. However, the synthetic P-C relations are in agreement with the empirical LMC $[3.6]-[4.5]$ P-C relation if a period range of $1.0<\log (P)<1.8$ is adopted when constructing the synthetic P-C relations. Observations of a large number of Cepheids with {\it JWST} may help in resolving the discrepancy of the IRAC band P-C relations.

\acknowledgments

The authors thank the referee for constructive comments to improve this manuscript. CCN thanks the funding from National Science Council (of Taiwan) under the contract NSC 98-2112-M-008-013-MY3. This work is based (in part) on observations made with the {\it Spitzer Space Telescope}, which is operated by the Jet Propulsion Laboratory, California Institute of Technology under a contract with NASA.


\begin{thebibliography}{}

\bibitem[Berdnikov et al(1996)]{ber96} Berdnikov, L. N., Vozyakova, O. V. \& Dambis, A. K., 1996, Astronomy Letters, 22, 838

\bibitem[Bono et al.(1999)]{bon99} Bono, G., Marconi, M., \& Stellingwerf, R.~F., 1999, \apjs, 122, 167 

\bibitem[Bono et al.(2000)]{bon00} Bono, G., Castellani, V., \& Marconi, M.\ 2000, \apj, 529, 293 

\bibitem[Bono et al.(2002)]{bon02} Bono, G., Castellani, V., \& Marconi, M.\ 2002, \apjl, 565, L83 

\bibitem[Bono et al.(2008)]{bon08} Bono, G., Caputo, F., Fiorentino, G., Marconi, M., \& Musella, I.\ 2008, \apj, 684, 102 

\bibitem[Bono et al.(2010)]{bon10} Bono, G., Caputo, F., Marconi, M., \& Musella, I.\ 2010, \apj, 715, 277 

\bibitem[Caputo et al.(2000)]{cap00} Caputo, F., Marconi, M. \& Musella, I., 2000, \aap, 354, 610

\bibitem[Caputo et al.(2002)]{cap02} Caputo, F., Marconi, M., \& Musella, I.\ 2002, \apj, 566, 833 

\bibitem[Caputo et al.(2005)]{cap05} Caputo, F., Bono, G., Fiorentino, G., Marconi, M., \& Musella, I.\ 2005, \apj, 629, 1021 

\bibitem[Fiorentino et al.(2002)]{fio02} Fiorentino, G., Caputo, F., Marconi, M. \& Musella, I., 2002, \apj, 576, 402

\bibitem[Fiorentino et al.(2007)]{fio07} Fiorentino, G., Marconi, M., Musella, I., \& Caputo, F.\ 2007, \aap, 476, 863 

\bibitem[Fouqu{\'e} et al.(2007)]{fou07} Fouqu{\'e}, P., et al., 2007, \aap, 476, 73 

\bibitem[Freedman et al.(2008)]{fre08} Freedman, W.~L., Madore, B.~F., Rigby, J., Persson, S.~E. \& Sturch, L., 2008, \apj, 679, 71 

\bibitem[Freedman \& Madore(2010)]{fre10} Freedman, W.~L. \& Madore, B.~F., 2010, Annual Reviews of Astronomy and Astrophysics, 48, 673

\bibitem[Freedman \& Madore(2011)]{fre11} Freedman, W.~L. \& Madore, B.~F., 2011, \apj, 734, 46

\bibitem[Freedman et al.(2011)]{fre11b} Freedman, W.~L., Madore, B.~F., Scowcroft, V., et al.\ 2011, \aj, 142, 192

\bibitem[Girardi et al.(2002)]{gir2002} Girardi, L., et al., 2002, \aap, 391, 195

\bibitem[Kennicutt et al.(1998)]{ken98} Kennicutt, R.~C., Jr., et al., 1998, \apj, 498, 181

\bibitem[Kutner et al.(2005)]{kut05} Kutner, M., Nachtsheim, C., Neter, J. \& Li, W., 2005, Applied Linear Statistical Models (fifth edition), McGraw-Hill, Singapore
 
\bibitem[Madore \& Freedman(1991)]{mad91} Madore, B. \& Freedman, W., 1991, \pasp, 103, 933

\bibitem[Madore et al.(2009)]{mad09} Madore, B.~F., Freedman, W.~L., Rigby, J., Persson, S.~E., Sturch, L. \& Mager, V., 2009, \apj, 695, 988 

\bibitem[Marconi et al.(2005)]{mar05} Marconi, M., Musella, I., \& Fiorentino, G., 2005, \apj, 632, 590

\bibitem[Marconi et al.(2010)]{marc10} Marconi, M., et al., 2010, \apj, 713, 615 

\bibitem[Marengo et al.(2010)]{mar10} Marengo, M., Evans, N.~R., Barmby, P., Bono, G., Welch, D.~L. \& Romaniello, M., 2010, \apj, 709, 120 

\bibitem[Neilson et al.(2010)]{nei10} Neilson, H.~R., Ngeow, C.-C., Kanbur, S.~M., \& Lester, J.~B.\ 2010, \apj, 716, 1136 

\bibitem[Ngeow \& Kanbur(2008)]{nge08} Ngeow, C.-C. \& Kanbur, S.~M., 2008, \apj, 679, 76 

\bibitem[Ngeow et al.(2009)]{nge09} Ngeow, C.-C., Kanbur, S.~M., Neilson, H.~R., Nanthakumar, A. \& Buonaccorsi, J., 2009, \apj, 693, 691

\bibitem[Ngeow \& Kanbur(2010)]{nge10} Ngeow, C.-C. \& Kanbur, S.~M., 2010, \apj, 720, 626

\bibitem[Ngeow et al.(2010)]{nge10b} Ngeow, C.-C., Ita, Y., Kanbur, S.~M., Neilson, H., Onaka, T. \& Kato, D., 2010, \mnras, 408, 983

\bibitem[Scowcroft et al.(2011)]{sco11} Scowcroft, V., Freedman, W., Madore, B.~F., Monson, A.~J., Persson, S.~E., Seibert, M., Rigby, J.~R., \& Sturch, L.\ 2011, \apj\ In-press (arXiv:1108.4672) 

\bibitem[Soszynski et al.(2008)]{sos08} Soszynski, I., et al., 2008, Acta Astron., 58, 163 

\end{thebibliography}
\end{document}